\documentclass[a4paper,11pt]{article}
\usepackage{ifpdf} 
\ifpdf
\pdfoutput=1
\fi

\usepackage{jcappub}
\usepackage{graphicx}
\graphicspath{{./}{figures/}}
\usepackage{dcolumn}
\usepackage{amssymb,amsmath,bm}
\usepackage{color}
\usepackage[dvipsnames]{xcolor}
\usepackage{xfrac}
\usepackage{aas_macros}
\usepackage{mathrsfs}
\usepackage{subcaption}
\usepackage{rotating}
\usepackage{chngcntr}
\usepackage[T1]{fontenc}
\usepackage{natbib}
\newcommand{\nv}{\hat{\bf n}}
\newcommand{\wtj}[6]{\left(\begin{array}{ccc} #1 & #2 & #3\\#4 & #5 & #6\end{array} \right)}

\newcommand{\bicep}{BICEP2/{\it Keck}}
\newcommand{\planck}{{\sl Planck}}

\newcommand{\oxford}{Department of Physics, University of Oxford, Denys Wilkinson Building, Keble Road, Oxford OX1 3RH, United Kingdom}
\newcommand{\ipmu}{Kavli Institute for the Physics and Mathematics of the Universe (Kavli IPMU, WPI), UTIAS, The University of Tokyo, Kashiwa, Chiba 277-8583, Japan}
\newcommand{\newcastle}{School of Mathematics, Statistics and Physics, Herschel Building, Newcastle University, Newcastle upon Tyne, NE1 7RU, UK}

\definecolor{internationalkleinblue}{rgb}{0.0, 0.18, 0.65}
\hypersetup{urlcolor=internationalkleinblue, linkcolor=internationalkleinblue, citecolor=internationalkleinblue}

\title{A minimal power-spectrum-based moment expansion for CMB B-mode searches}
\author[1,2]{S. Azzoni,}
\author[1]{M. H. Abitbol,}
\author[1]{D. Alonso,}
\author[1,3]{A. Gough,}
\author[2]{N. Katayama,}
\author[2]{T. Matsumura}

\affiliation[1]{\oxford}
\affiliation[2]{\ipmu}
\affiliation[3]{\newcastle}
\emailAdd{susanna.azzoni@physics.ox.ac.uk}

\abstract{The characterization and modeling of polarized foregrounds has become a critical issue in the quest for primordial $B$-modes. A typical method to proceed is to factorize and parametrize the spectral properties of foregrounds and their scale dependence (i.e. assuming that foreground spectra are well described everywhere by their sky average). Since in reality foreground properties vary across the Galaxy, this assumption leads to inaccuracies in the model that manifest themselves as biases in the final cosmological parameters (in this case the tensor-to-scalar ratio $r$). This is particularly relevant for surveys over large fractions of the sky, such as the Simons Observatory (SO), where the spectra should be modeled over a distribution of parameter values. Here we propose a method based on the existing ``moment expansion'' approach to address this issue in a power-spectrum-based analysis that is directly applicable in ground-based multi-frequency data. Additionally, the method uses only a small set of parameters with simple physical interpretation, minimizing the impact of foreground uncertainties on the final $B$-mode constraints. We validate the method using SO-like simulated observations, recovering an unbiased estimate of the tensor-to-scalar ratio $r$ with standard deviation $\sigma(r)\simeq0.003$, compatible with official forecasts. When applying the method to the public \bicep{} data, we find an upper bound $r<0.06$ ($95\%\,{\rm C.L.}$), compatible with the result found by \bicep{} when parametrizing spectral index variations through a scale-independent frequency decorrelation parameter. We also discuss the formal similarities between the power spectrum-based moment expansion and methods used in the analysis of CMB lensing.}

\begin{document}
\maketitle
\flushbottom

  \section{Introduction}\label{sec:intro}
    The Cosmic Microwave Background (CMB) primordial, parity-odd ``$B$-mode'' polarization signal encodes crucial information on the physics of the early Universe \cite{1997PhRvL..78.2058K, 1997PhRvL..78.2054S}. Within the inflationary paradigm, tensor metric perturbations are generated during the primordial exponential expansion of the Universe. These then give rise to $B$-mode polarization in the CMB photons via Thomson scattering. The primordial $B$-mode amplitude is usually parametrized in terms of $r$, the ratio between the power spectra of tensor and scalar primordial perturbations, which can be directly linked to the energy scale at which inflation occurred \cite{1997PhRvL..78.1861L}. If detected with a sufficiently large amplitude, this faint cosmological signal would allow us to exclude several families of inflationary models, as well as non-inflationary alternatives. However, there is no firm prediction for the value of $r$ in inflationary models, as inflation can in principle create tensor perturbations with arbitrarily small amplitudes. A large family of models, such as Higgs or $R^2$ inflation predicts values of the order $r\sim1/N^2\sim0.001$ \cite{1979JETPL..30..682S,2008PhLB..659..703B}, where $N$ is the number of $e$-folds inflation lasts. Current best upper limits are set by the \bicep{} collaboration at $r < 0.07$ (95$\%$ CL) \cite{2016PhRvL.116c1302B}, and $r < 0.044$ in the latest combination with \planck{} \citep{2020arXiv201001139T}. However, the forthcoming generation of CMB polarization observatories have been designed to achieve sensitivities able to reach statistical uncertainties equivalent to $\sigma(r)\sim10^{-3}$ on the tensor-to-scalar ratio  \cite{2014SPIE.9153E..1PB,2014JLTP..176..733M,2016arXiv161002743A,2018SPIE10708E..07H,2018PhRvL.121v1301B,2018SPIE10698E..4FS,2019JCAP...02..056A}.
    
    The primordial $B$-mode signal, however, is extremely faint compared to other sources of $B$-modes. On the one hand, gravitational lensing by the intervening large-scale structure generates a $B$-mode contribution which, on small scales, is similar to a white-noise component with amplitude $\sigma_N=5\,\mu K\,{\rm arcmin}$ \cite{2002PhRvL..89a1303K}. On the other hand, the most important astrophysical source is the contamination from Galactic polarized foregrounds \cite{2014PhRvL.112x1101B}. In particular, polarized synchrotron dominates the sky emission at low frequencies ($\nu\lesssim40\,{\rm GHz}$), while thermal dust emission is more relevant at higher frequencies ($\nu\gtrsim150\,{\rm GHz}$). The combination of both, in any case, dominates over the CMB $B$-mode signal, including the lensing contribution, over the whole frequency range on degree scales. The separation of the multi-frequency data into different components is therefore a crucial part in the analysis of CMB $B$-mode data \cite{2014PhRvL.112x1101B,2015PhRvL.114j1301B,2016MNRAS.458.2032R,2016JCAP...03..052E,2017PhRvD..95d3504A,2019arXiv191002608A}. Manifestly optimal map-based component separation methods have been designed and applied to existing satellite datasets \citep{2004ApJS..155..227E,2008ApJ...676...10E,2009ApJ...701.1804D,2011MNRAS.418..467R,2020A&A...641A...4P,2020arXiv200608628R}. However, their implementation on high-resolution datasets is computationally challenging due to the larger number of modes. Although approximate parametric map-level methods exist (e.g. \cite{2009MNRAS.392..216S,2016PhRvD..94h3526S,2020MNRAS.496.4383G}), that can deal with large numbers of pixels more efficiently, their implementation on ground-based data is further complicated by the presence of complex filtering and inhomogeneous non-white noise introducing non-trivial correlations between pixels. Partly for this reason, multi-frequency $C_\ell$-based approaches, where the signal is modeled directly at the level of the cross-frequency power spectra, and where some of these complications are easier to deal with (e.g. through the use of transfer functions \citep{2002ApJ...567....2H}), have been developed and used for ground-based data \cite{2013JCAP...07..025D,2015PhRvL.114j1301B,2018PhRvL.121v1301B,2020arXiv200707289C}.

    The main drawback of these methods, in their simplest incarnation, is the difficulty to account for the expected spatial variability of foreground spectral properties. Although the main effects of this spatial variability, in the form of frequency decorrelation \citep{2017A&A...599A..51P,2018PhRvD..97d3522S}, can be effectively taken into account in specific cases (e.g. uncorrelated spectral index variations \citep{2017A&A...603A..62V}), developing a framework to account for this variability in a general scenario will be useful in the analysis of data from ongoing and future ground-based experiments given their higher sensitivity. In this paper, we will make use of the so-called moment expansion formalism, introduced in \cite{1998ApJ...502....1T,2004ApJ...608..622Z,2005ApJ...625..575S,2017MNRAS.472.1195C}, to derive the simplest extension to the standard power spectrum-level parametrizations of foreground spatial variability, and quantify the ability of this method to obtain unbiased constraints on the tensor-to-scalar ratio from existing data and upcoming experiments. This problem was already addressed by \cite{2019arXiv191209567M}, where the most general moment expansion was presented and applied to \planck{} data. In this work, we instead study the simplest version of this expansion, and evaluate its performance for next-generation ground-based CMB data using foreground simulations with varying levels of realism.

    This paper is structured as follows. Section \ref{sec:theory} describes the basic elements of power-spectrum-based component separation methods, the main effects of spatially-varying foreground spectral properties, and introduce the moment expansion method we will use to account for those. Section \ref{sec:sims} describes the ingredients of the synthetic sky simulations used to validate the method. This validation is described in detail in Section \ref{sec:results}, which presents the limits of applicability of the leading-order moment expansion and its performance on simulations with varying degrees of foreground complexity. After validating the method, using simulated observations mimicking the expected performance of the Simons Observatory (SO), we apply it to the public data from the \bicep{} collaboration and obtain constraints on $r$ marginalized over foreground spatial variations. We summarize and discuss our results in Section \ref{sec:conclusion}. Appendix \ref{app:corr} presents a more accurate, non-perturbative calculation of the impact of foreground spatial variations using a formalism similar to that used to estimate the effects of gravitational lensing on the primary CMB power spectrum.

  \section{Formalism}\label{sec:theory}
    \subsection{Sky model}\label{ssec:theory.map}
      We model the polarized sky signal at position $\nv$ and frequency $\nu$, ${\bf m}_\nu(\nv)\equiv(Q_\nu(\nv),U_\nu(\nv))$, as a sum of components of the form
      \begin{equation}\label{eq:map_1}
        {\bf m}_\nu(\nv)=\sum_c {\bf T}_c(\nv)\,S_\nu^c(\vec{\beta}_c(\nv)),
      \end{equation}
      where ${\bf T}_c(\nv)$ is the amplitude of component $c$ at a pivot frequency $\nu_0^c$, and $S_\nu^c$ is its frequency spectrum (normalized to $S_{\nu_0}^c=1$). $\vec{\beta}_c(\nv)$ is a set of parameters describing the spectrum, which can vary as a function of sky position.
  
      We will consider three components:
      \begin{itemize}
        \item {\bf CMB}: in antenna temperature units, the spectrum of CMB temperature anisotropies is
        \begin{equation}
          S_\nu^{\rm CMB}=e^x\left(\frac{x}{e^x-1}\right)^2,\hspace{12pt}x=\frac{h\nu}{k_B \Theta_{\rm CMB}},
        \end{equation}
        where $h$ is the Planck constant, $k_B$ is the Boltzmann constant, and $\Theta_{\rm CMB}=2.7255\,{\rm K}$ is the CMB monopole temperature~\cite{fixsen2009}. The CMB spectrum is isotropic and is not normalized at any pivot frequency.
        \item {\bf Thermal dust:} dust grains in the interstellar medium are heated by stellar radiation, producing emission on microwave frequencies. The alignment of elongated dust grains with the Galactic magnetic field (GMF) produces a linear polarization perpendicular to both the magnetic field and the direction of propagation, making dust the most relevant foreground for $B$-mode searches on frequencies $\nu\gtrsim150\,{\rm GHz}$. Thermal dust emission is well-characterized by a modified black-body (MBB) spectrum of the form \citep{2015A&A...576A.107P}
        \begin{equation}\label{eq:mbb}
          S_\nu^{\rm D} = \left(\frac{\nu}{\nu^{\rm D}_0}\right)^{\beta_{\rm D}}\frac{B_\nu(\Theta_{\rm D})}{B_{\nu^{\rm D}_0}(\Theta_{\rm D})},
        \end{equation}
        where $\beta_{\rm D}$ and $\Theta_{\rm D}$ are the dust spectral index and temperature, and
        \begin{equation}
          B_\nu(\Theta) = \frac{2h\nu^3}{c^2}\left[\exp\left(\frac{h\nu}{k\Theta}\right)-1\right]^{-1}
        \end{equation}
        is the Planck black-body spectrum. We will consider spatial variations in $\beta_{\rm D}$, which takes values $\beta_{\rm D}\sim1.6$. The restricted frequency range available to most ground-based experiments, including the SO ($\nu\lesssim280\,{\rm GHz}$), makes $B$-mode studies almost insensitive to the value of $\Theta_{\rm D}$, and therefore we fix it to $\Theta_{\rm D}=19.6\,{\rm K}$ here. Note that departures from a pure MBB law are anticipated by dust models (e.g. \cite{2013ApJ...765..159D,2017ApJ...834..134H}).
        \item {\bf Synchrotron:} Galactic synchrotron emission is caused by the interaction of high-energy cosmic ray electrons with the GMF \citep{1986rpa..book.....R}. Synchrotron is strongly polarized, and is characterized by a smooth power-law spectrum tracing the energy distribution of cosmic ray electrons. The synchrotron spectrum used here is therefore
        \begin{equation}
          S_\nu^{\rm S}=\left(\frac{\nu}{\nu_0^{\rm S}}\right)^{\beta_{\rm S}},
        \end{equation}
        where $\beta_{\rm S}$ is the synchrotron spectral index, which takes values $\beta_{\rm S}\sim-3$.
      \end{itemize}
      The spatially-varying degrees of freedom of the sky model are therefore the amplitudes of the three components $({\bf T}_{\rm CMB}(\nv),\,{\bf T}_{\rm D}(\nv),\,{\bf T}_{\rm S}(\nv))$ and the two spectral indices $(\beta_{\rm D}(\nv),\,\beta_{\rm S}(\nv))$, where we will use the notation `D' to indicate dust parameters and `S' for synchrotron.

      A full characterization of these variables would ideally require a pixel-based component separation approach (e.g. \cite{2008ApJ...676...10E,2009ApJ...701.1804D,2016PhRvD..94h3526S,2017PhRvD..95d3504A}) where they are constrained in each pixel individually. Unfortunately this approach would significantly degrade the uncertainties in the final $B$-mode constraints, due to the limited frequency coverage of ground-based experiments \citep{2016PhRvD..94h3526S,2017PhRvD..95d3504A} (e.g. 6 bands in the case of SO). Although ``pooling'' approaches have been proposed where single spectral indices are constrained in larger sky regions to reduce the number of degrees of freedom, the degradation of final constraints can still be significant \cite{2020MNRAS.496.4383G}, and the resulting component-separated amplitude maps often retain features associated with the choice of pooling. Moreover, especially in the case of ground-based data, map-based analyses can be extremely complex, given the various filtering operations that the final frequency maps are subjected to, which can lead to important non-linear biases in the component-separated maps \citep{2014A&A...571A..15P}. On the other hand there exist well-tested methods to account for these effects when computing power spectra between different frequency maps (e.g. the use of cross-split correlations to avoid imperfections in modeling the noise bias, or the use of transfer functions or observation matrices to correct for the impact of map-level filtering operations \citep{2016ApJ...825...66B,2019arXiv191002608A,2020arXiv200707289C}). For this reason, most cosmological analyses from ground-based data have used a ``multi-frequency'' likelihood where the analysis is based on modeling the full set of cross-frequency power spectra \cite{2016PhRvL.116c1302B,2018PhRvL.121v1301B,2019arXiv191002608A}. This will motivate us propagating the map-based model described here to the power spectrum level in Section \ref{ssec:theory.cl}.
  
      We can make further progress by assuming that the spatial variations in the spectral parameters $\beta_c^i$ across the mapped footprint are small, and using the ``moments-based'' expansion \cite{2017MNRAS.472.1195C}. Expanding $\beta_c^i=\bar{\beta}_c^i+\delta\beta_c^i(\nv)$, where $\bar{\beta}_c^i$ is the mean of $\beta_c^i$ across the map, and $\delta\beta_c^i$ describes its spatial fluctuations, we can expand Eq. \ref{eq:map_1} to second order in $\delta\beta_c^i$ as
      \begin{equation}\label{eq:map_2}
        {\bf m}_\nu(\nv)=\sum_c\left[
        {\bf T}_c(\nv)\bar{S}^c_\nu+
        {\bf T}_c(\nv)\,\delta\beta_c^i(\nv)\,\partial_i\bar{S}^c_\nu+
        \frac{1}{2!}{\bf T}_c(\nv)\,\delta\beta_c^i(\nv)\,\delta\beta_c^j(\nv)\,\partial_i\partial_j\bar{S}^c_\nu+\mathcal{O}(\delta\beta^3)\right],
      \end{equation}
      where we have used the shorthand
      \begin{equation}
        \bar{S}^c_\nu=S^c_\nu(\bar{\beta}),\hspace{12pt}\partial_i\bar{S}^c_\nu\equiv\left.\frac{\partial S_\nu^c}{\partial\beta_c^i}\right|_{\beta=\bar{\beta}},\hspace{12pt}\partial_i\partial_j\bar{S}^c_\nu\equiv\left.\frac{\partial^2 S_\nu^c}{\partial\beta_c^i\partial\beta_c^j}\right|_{\beta=\bar{\beta}}.
      \end{equation}
      It is worth noting that the derivatives of the spectrum take a specific form in the case of spectral indices, which enter the spectrum as $(\nu/\nu_0)^\beta$:
      \begin{equation}
        \frac{\partial^n S_\nu^c}{\partial\beta^n}=\left[\log\left(\frac{\nu}{\nu_0^c}\right)\right]^n\,S_\nu^c.
      \end{equation}
      In Eq. \ref{eq:map_1} and in what follows we have used Einstein's notation with respect to summation over repeated indices identifying spectral parameters ($i$, $j$ etc.).

    \subsection{$C_{\ell}$-based cleaning method}\label{ssec:theory.cl}
      Let us now propagate the formalism above to the power spectrum between two frequency maps
      \begin{equation}\label{eq:cldef}
        \langle a^\nu_{\ell m}a^{\nu'*}_{\ell' m'}\rangle=\delta_{\ell\ell'}\delta_{mm'}C_\ell^{\nu\nu'},
      \end{equation}
      where $a_{\ell m}^\nu$ are the spherical harmonic coefficients associated with map $m_\nu(\nv)$, which we have assumed is statistically isotropic.  Note that, for simplicity, we will assume that all fields involved ($m_\nu(\nv),\,T_c(\nv)$) are spin-0 fields, even though our main application, the polarization of the CMB, is a spin-2 field. The generalization of our main result to fields of arbitrary spin is presented in Appendix \ref{app:spin2}, where we justify that neglecting the spin-2 nature is a good approximation in our analysis\footnote{This is further reinforced by the good agreement of our formalism with Gaussian simulations presented in Section \ref{ssec:res.gaussian}.}. We will thus apply these results directly to the pseudo-scalar $B$-mode component.
    
      Using Eq. \ref{eq:map_2}, the power spectrum $C_\ell^{\nu\nu'}$ is, up to second order in $\delta\beta$, given by:
      \begin{align}\label{eq:c_ell_1}
        C_\ell^{\nu\nu'}=C_\ell^{\nu\nu'}|_{0\times0}+C_\ell^{\nu\nu'}|_{0\times1}+C_\ell^{\nu\nu'}|_{1\times1}+C_\ell^{\nu\nu'}|_{0\times2},
      \end{align}
      where
      \begin{align}\label{eq:c_ell_00}
        &C_\ell^{\nu\nu'}|_{0\times0}\equiv \sum_{cc'}\bar{S}^c_\nu\,\bar{S}_{\nu'}^{c'}\,C_\ell(T_c,T_{c'}),\\\label{eq:c_ell_01}
        &C_\ell^{\nu\nu'}|_{0\times1}\equiv \sum_{cc'}\partial_i\bar{S}^c_\nu\,\bar{S}_{\nu'}^{c'}\,C_\ell(T_c\delta\beta_c^i,T_{c'})+(\nu\leftrightarrow\nu'),\\\label{eq:c_ell_11}
        &C_\ell^{\nu\nu'}|_{1\times1}\equiv \sum_{cc'}\partial_i\bar{S}^c_\nu\,\partial_j\bar{S}_{\nu'}^{c'}\,C_\ell(T_c\delta\beta_c^i,T_{c'}\delta\beta_{c'}^j),\\\label{eq:c_ell_02}
        &C_\ell^{\nu\nu'}|_{0\times2}\equiv \frac{1}{2}\sum_{cc'}\partial_i\partial_j\bar{S}^c_\nu\,\bar{S}_{\nu'}^{c'}\,C_\ell(T_c\delta\beta_c^i\delta\beta_c^j,T_{c'})+(\nu\leftrightarrow \nu'),
      \end{align}
      where $C_\ell(a,b)$ denotes the power spectrum between fields $a$ and $b$, and $(\nu\leftrightarrow\nu')$ implies the same term swapping the roles of $\nu$ and $\nu'$.
      
      The $0\times0$ term, which is exact in the absence of spatial variations of spectral indices, is the basis for the fiducial power-spectrum level cleaning methods used by CMB collaborations \citep{2016PhRvL.116c1302B,2018PhRvL.121v1301B}, and it involves the modeling of $N_c(N_c+1)/2$ power spectra for $N_c$ different components. Assuming one single free spectral parameter for each component, the $0\times1$, $1\times1$ and $0\times2$ terms would in general imply modeling an additional $N_c^2$, $N_c(N_c+1)/2$ and $N_c^2$ different power spectra respectively (for a total of $N_c(3N_c+1)$ spectra). These individual contributions are associated to distinct spectral responses in terms of the combination of spectrum derivatives that accompany them, and it should therefore be possible to separate all contributions cleanly given a large enough number of frequency channels $N_\nu$ (leading to $N_\nu(N_\nu+1)/2$ distinct cross-frequency correlations). For example, on a bandpower-by-bandpower basis, with two foreground sources, associated with a single spectral parameter each, as well as the CMB, this approach would involve modeling 15 different cross-spectra (compared with 4 spectra if the higher-order terms are neglected). An experiment with 6 frequency channels would be able to measure 21 different cross-frequency spectra, and should therefore be able to separate the 15 different contributions, albeit at a significant cost in statistical uncertainties \citep{2019arXiv191209567M}. In the presence of more than one free spectral parameter per component, the number of independent spectra to model would increase rapidly.
      
      With the objective of ameliorating the complexity of this model, we will make the following three simplifying assumptions:
      \begin{enumerate}
        \item In the $1\times1$ and $0\times2$ terms, we will assume that different components are uncorrelated, i.e. terms like $C_\ell(T_{\rm S}\delta\beta_{\rm S},T_{\rm D}\delta\beta_{\rm D})$ are zero. We know that this assumption is wrong at some level. Polarized dust and synchrotron are associated to the same GMF, and there is evidence that they are correlated on large scales. We will indeed account for this correlation in the $0\times0$ term but ignore it in the higher-order ones, effectively treating the cross-component correlation coefficient as another perturbative parameter. 
        \item We will assume Gaussian statistics for all fields involved. This automatically cancels the $0\times1$ term, which only contains 3-point functions. Furthermore this allows us to express all four-point functions in Eqs. \ref{eq:c_ell_11} and \ref{eq:c_ell_02} as products of two-point functions. Galactic foregrounds are well-known to be non-Gaussian, although the assumption of Gaussianity may be a better approximation in polarization than intensity \citep{2017A&A...603A..62V}. Nevertheless, the rationale is the same as before, treating the non-Gaussian foreground terms as higher-order in the perturbative expansion.
        \item We will assume that amplitudes $T_c$ and spectral index fluctuations $\delta\beta_c$ are uncorrelated, therefore ignoring terms of the form $C_\ell(T_c,\delta\beta_c)$. This is also not generally true, since variations in spectral indices are likely to trace the same structures (e.g. dust filaments) that generate the foreground signals. As before, we ignore these correlations, treating the associated correlation coefficients as additional perturbative parameters that would make them higher-order in the expansion.
      \end{enumerate}
      
      These assumptions yield the simplest possible description of the multi-frequency power spectrum at leading order in the spatial variation of the foreground spectral parameters. For the specific model used here (synchrotron and dust with free spectral indices), the three surviving contributions to Eq. \ref{eq:c_ell_1} read:
      \begin{align}\label{eq:mom00}
        &C_\ell^{\nu\nu'}|_{0\times0}=\bar{S}^{\rm D}_\nu\bar{S}^{\rm D}_{\nu'}\,C_\ell^{\rm DD}+\bar{S}^{\rm S}_\nu\bar{S}^{\rm S}_{\nu'}\,C_\ell^{\rm SS}+\left(\bar{S}^{\rm D}_\nu\bar{S}^{\rm S}_{\nu'}+\bar{S}^{\rm S}_\nu\bar{S}^{\rm D}_{\nu'}\right)C^{\rm SD}_\ell,\\\label{eq:mom11}
        &C_\ell^{\nu\nu'}|_{1\times1}=\sum_{c\in\{{\rm D},{\rm S}\}}\partial_\beta\bar{S}^c_\nu\,\partial_\beta\bar{S}^c_{\nu'}\sum_{\ell_1\ell_2}\frac{(2\ell_1+1)(2\ell_2+1)}{4\pi}\wtj{\ell}{\ell_1}{\ell_2}{0}{0}{0}^2C_{\ell_1}^{cc}\,C_{\ell_2}^{\beta_c},\\\label{eq:mom02}
        &C_\ell^{\nu\nu'}|_{0\times2}=\sum_{c\in\{{\rm D},{\rm S}\}}\frac{1}{2}\left[\bar{S}^c_\nu\,\partial^2_\beta\bar{S}^c_{\nu'}+\bar{S}^c_{\nu'}\,\partial^2_\beta\bar{S}^c_\nu\right]C_\ell^{cc}\sigma_{\beta_c}^2,
      \end{align}
      where we have used the shorthand
      \begin{equation}
        C_\ell^{cc}\equiv C_\ell(T_c,T_c),\hspace{12pt}
        C_\ell^{\beta_c}\equiv C_\ell(\beta_c,\beta_c),\hspace{12pt}
        \sigma_{\beta_c}^2\equiv\sum_\ell \frac{2\ell+1}{4\pi}C_\ell^{\beta_c}.
      \end{equation}
      At the cost of generality, the method is therefore significantly simpler, requiring only the modeling of two additional power spectra, $C_\ell^{\beta_{\rm D}}$ and $C_\ell^{\beta_{\rm S}}$. We have explored the impact of the simplifying assumptions used here through the analysis of idealized and realistic simulations, as well as precursor data, as described in Section~\ref{sec:results}.
      
      We parametrize the different ingredients described above by extending the model used by \cite{2018PhRvL.121v1301B} for the $0\times0$ contribution. The amplitude power spectra are modeled as power laws of the form
      \begin{align}\label{eq:cl_amp}
        \frac{\ell(\ell+1)}{2\pi}C_\ell^{cc}=A_c\left(\frac{\ell}{\ell_0}\right)^{\alpha_c},\hspace{12pt}
      \end{align}
      with the dust-synchrotron cross-correlation parametrized through a scale-independent correlation coefficient: $C_\ell^{\rm SD}=\epsilon_{\rm DS}\sqrt{C_\ell^{\rm DD}C_\ell^{\rm SS}}$. Likewise, the power spectrum of $\delta\beta_c$ is parametrized as
      \begin{align}\label{eq:cl_beta}
        C_\ell^{\beta_c}=B_c\left(\frac{\ell}{\ell_0}\right)^{\gamma_c},
      \end{align}
      with $\ell_0=80$ in all cases. Finally, the CMB $B$-mode power spectrum is parametrized as
      \begin{equation}\label{eq:cl_cmb_bb}
       C_{\ell}^{\rm CMB} = A_{\rm lens}\,C^{\rm lens}_\ell+r\,\left.C^{\rm tens}_\ell\right|_{\rm{r = 1}},
      \end{equation}
      where $C^{\rm lens}_\ell$ and $C^{\rm tens}_\ell|_{\rm{r=1}}$ are templates for the $B$-mode power spectrum caused by gravitational lensing and by primordial tensor fluctuations with tensor-to-scalar ratio $r=1$ respectively. The model therefore has 13 free parameters:
      \begin{equation}
       \{r,A_{\rm lens},A_{\rm D},\alpha_{\rm D},\beta_{\rm D},B_{\rm D},\gamma_{\rm D},A_{\rm S},\alpha_{\rm S},\beta_{\rm S},B_{\rm S},\gamma_{\rm S},\epsilon_{\rm SD}\}.
      \end{equation}

    \subsection{Power spectrum likelihood}\label{ssec:theory.likelihood}
      To derive constraints on the free parameters of the model, we will use a multi-frequency power spectrum likelihood. In this case, the data vector is the full matrix of cross-frequency power spectra $C_\ell^{\nu\nu'}$, with the corresponding theory prediction described in the previous section. On large scales, the small number of available modes invalidates the central limit theorem and, as quadratic functions of Gaussian fields, power spectra exhibit non-Gaussian features in their likelihoods. To account for this effect, we use the non-Gaussian likelihood developed by \cite{2008PhRvD..77j3013H} (HL hereon). This likelihood requires an estimate of the covariance matrix of the full set of power spectra. In order to accurately account for the effects of incomplete sky coverage and $EB$ leakage, we estimate this covariance matrix from a set of 500 Gaussian simulations, generated as described in Section~\ref{ssec:sims.gaussian}. The HL likelihood additionally requires an estimate of the fiducial power spectra, as well as the noise power spectrum. We produce the former from the fiducial set of parameters used to generate the simulations (see Section~\ref{ssec:sims.gaussian}), and the latter by averaging the power spectra of the 500 noise realizations generated for these simulations. Atmospheric noise and various systematics will likely limit the largest scales that can be reliably used by ground-based experiments, and therefore, in addition to the use of the realistic noise curves described in Section~\ref{ssec:sims.noise}, we use the restricted scale range $30\leq\ell\leq300$ where $B$-mode signal from the recombination bump is concentrated \citep{2018PhRvL.121v1301B}.
      \begin{table}[tbp]
        \centering
        \begin{tabular}{|l|c|c|}
        \hline
        Parameter & Prior & Bounds\\
        \hline
        $r$ & Top-hat & [-1, 1] \\
        $A_{\rm lens}$ & Top-hat & [0, 2]\\
        $A_D$ & Top-hat & [0, $\infty$] \\
        $\alpha_D$ & Top-hat & [-1, 0] \\
        $\beta_D$ & Gaussian &  1.6 $\pm$ 0.5\\
        $\gamma_D$ & Top-hat & [-6, -2] \\
        $B_D$ & Top-hat & [0, 10] \\
        $A_S$ & Top-hat & [0, $\infty$] \\
        $\alpha_S$ & Top-hat & [-1, 0] \\
        $\beta_S$ & Gaussian &  -3 $\pm$ 0.6 \\
        $B_S$ & Top-hat & [0, 10] \\
        $\gamma_S$ & Top-hat & [-6, -2] \\
        $\epsilon_{SD}$ & Top-hat & [-1, 1] \\
        \hline
        \end{tabular}
        \caption{Summary of the priors used in the analysis. Note that all amplitude parameters (except $r$) have physically-motivated positivity priors.} \label{tab:priors}
      \end{table}
      
      The posterior distribution is given by the product of this likelihood and a set of priors. The priors have been chosen to be wide enough that the parameters are constrained by the data in most cases. They are summarized in Table \ref{tab:priors}. Note that we impose a physically motivated priors on all power spectrum amplitudes ($A_c$, $B_c$) forcing them to be positive. An exception to this is $r$, which we allow to be negative in order to detect possible negative biases. We sample this posterior distribution using the affine-invariant Monte-Carlo Markov chain ensemble sampler {\tt emcee} \cite{2013PASP..125..306F}. The chains were started around the maximum likelihood point, found via Powell's minimization scheme \cite{powell1964efficient} as implemented in {\tt scipy} \cite{2020SciPy-NMeth}.

  \section{Simulations}\label{sec:sims}
    In order to test the validity of the moment expansion method described in the previous section, we test it on a suite of sky simulations. These simulations include the most relevant sky components, as discussed in Section~\ref{ssec:theory.map}, with varying degrees of realism in order to explore the impact of the assumptions of the method regarding foreground properties on its performance. The simulations also incorporate the contribution from instrumental noise and limited sky coverage for a SO-like experiment, as described in Section~\ref{ssec:sims.noise} .
  
    In all simulations, the CMB contribution was generated as a Gaussian random field drawn from the power spectrum in Eq.~\ref{eq:cl_cmb_bb}. We use fiducial values $(A_{\rm lens}=1,\,r=0)$ unless otherwise stated.

    \subsection{Gaussian foreground simulations}\label{ssec:sims.gaussian}
      We generate a large suite of ``Gaussian'' simulated skies. For these, we simulate sky maps for the amplitude and spectral index variation maps (${\bf T}_c(\nv)$ and $\delta\beta_c(\nv)$) as Gaussian random fields governed by power spectra following the power-law models in Eqs. \ref{eq:cl_amp} and \ref{eq:cl_beta}. We then add the mean spectral indices $\bar{\beta}_c$ to $\delta\beta_c(\nv)$ and use the Python Sky Model software ({\tt PySM}, \cite{thorne2017python}) to generate observed sky maps in the six SO frequency channels (see Section~\ref{ssec:sims.noise}). These maps are generated using the {\tt HEALPix} pixelization scheme with resolution parameter $N_{\rm side}=256$. The pixel resolution ($\delta\theta\sim0.2^\circ$) is enough to cover the $\ell$ range relevant for our analysis. Amplitude and spectral index maps were generated as uncorrelated Gaussian fields.
      
      The aim of these Gaussian simulations is twofold. First, by using the exact same model assumed by the cleaning method (Gaussian fields, uncorrelated indices and amplitudes), we can test the validity of the leading-order expansion in $\delta\beta_c$ for different levels of spectral index variation, and compare it with the full, non-perturbative result. Secondly, the HL likelihood used here (see Section \ref{ssec:theory.likelihood}) requires an estimate of the power spectrum covariance. In order to fully incorporate the effects of inhomogeneous noise, mode-coupling and $E/B$ mixing in the covariance matrix, we use these Gaussian simulations to compute it.
      
      To estimate covariance matrices, we generate two suites of 500 Gaussian simulations. Both suites were generated with constant spectral indices ($B_{\rm D}=B_{\rm S}=0$) but with different values for the amplitude power spectrum parameters. The first suite used the best-fit foreground parameters found by the \bicep{} collaboration \cite{2018PhRvL.121v1301B}:
      \begin{align}\nonumber
        &A_{\rm D}=5\,\mu{\rm K}^2,\hspace{12pt}\alpha_{\rm D}=-0.42,\hspace{12pt}
        A_{\rm S}=2\,\mu{\rm K}^2,\hspace{12pt}\alpha_{\rm S}=-0.6,
      \end{align}
      and was used in the analysis of the method's performance as a function of spectral index variation amplitude. The second suite used the foreground parameters that best fit the dust and synchrotron template maps used in the realistic set of simulations described in Section~\ref{ssec:sims.pysm}, and were used to both validate the method against realistic simulations as described in Section \ref{ssec:res.pysm}, and in the simulation challenge described in Section \ref{ssec:res.challenge}. The corresponding parameter values are
      \begin{align}\nonumber
        &A_{\rm D}=28\,\mu{\rm K}^2,\hspace{12pt} \alpha_{\rm D}=-0.16,\hspace{12pt} A_{\rm S}=1.6\,\mu{\rm K}^2,\hspace{12pt} \alpha_{\rm S}=-0.93.
      \end{align}
      All the Gaussian simulations presented here used the same values for the constant spectral indices ($\beta_{\rm D}=1.6$, $\beta_{\rm S}=-3$) and the same pivot frequencies ($\nu_0^{\rm D}=353\,{\rm GHz}$ and $\nu_0^{\rm S}=23\,{\rm GHz}$). The dust-synchrotron correlation coefficient was set to $\epsilon_{\rm SD}=0$.
      
      To study the performance of the moment expansion method for different levels of spectral index variation, we generate a number of additional simulations with spectral index maps generated as Gaussian realizations of $C_\ell^{\beta_c}$ in Eq.~\ref{eq:cl_beta} with varying values for the amplitude $B_c$. Instead of varying $B_c$ directly, we generate maps of $\delta\beta_c(\nv)$ with an arbitrary amplitude and then renormalize them to enforce a given per-pixel standard deviation $\sigma({\beta_c})\equiv\sqrt{\langle\delta\beta_c^2\rangle}$. Therefore our results will be presented in terms of $\sigma({\beta_c})$ as a more meaningful parameter, rather than $B_c$. Unless otherwise stated, we fix the spectral tilt $\gamma_c$ to the arbitrary values $(\gamma_{\rm D},\gamma_{\rm S})=(-3.5, -2.5)$. Note that we will also study the impact of the value of $\gamma_c$ on the results, since this parameter regulates the distribution of spectral index fluctuations on different scales.

    \subsection{Realistic foreground simulations}\label{ssec:sims.pysm}
      \begin{figure}[h!t]
        \centering
        \includegraphics[width=0.9\textwidth]{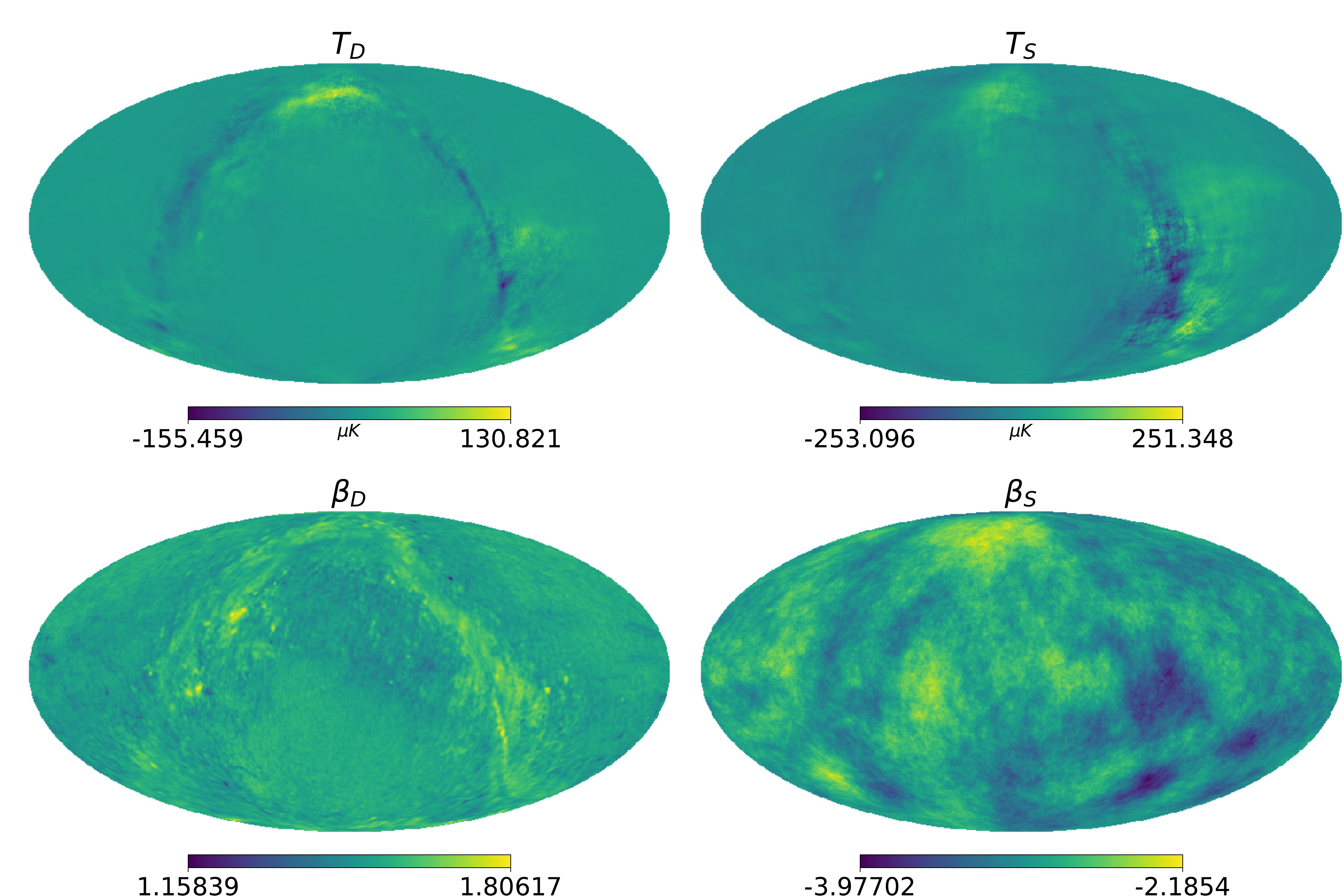}
        \caption{Dust and synchrotron polarized amplitude maps (\textit{top row}) and spectral index maps (\textit{bottom row}).}
        \label{fig:amps_and_betas}
      \end{figure}
      In order to test the validity of the assumptions adopted by our simplified moment expansion (Gaussianity, independence of spectral index and amplitudes, and between components), we produce an additional set of simulations with higher level of realism.
      
      To do so, we use the foreground templates provided by {\tt PySM}. Specifically, we use the dust amplitude map and spectral index map assumed by the {\tt d1} model, as well as the synchrotron amplitude map included in the {\tt s1} model. The synchrotron spectral index map provided with {\tt s1} was originally derived from a combination of the Haslam 408 GHz map \cite{haslam1981408, haslam1982408} and the WMAP 23 GHz map \cite{dunkley2009prospects, 2013ApJS..208...20B}. The map presents very mild fluctuations in $\beta_{\rm S}$, and only on very large scales. The level of variation and the overall value of $\beta_{\rm S}$ in this model has been shown by \cite{2018A&A...618A.166K} to be too low using data from the SPASS experiment. To increase the complexity of the synchrotron contribution we instead generate an alternative spectral index map generated by re-scaling the {\tt s1} map and extending it to smaller scales using a power-law spectral index power spectrum $C_\ell^{\beta_{\rm S}}$ matching the measurements of \cite{2018A&A...618A.166K}. The dust spectral index in {\tt d1} corresponds to the estimate of $\beta_{\rm D}(\nv)$ from the \planck{} data using the {\tt Commander} component separation code \citep{2016A&A...594A..10P}.

      Figure~\ref{fig:amps_and_betas} shows the polarized amplitude and spectral index maps used in these simulations. The rms fluctuation around the mean of the spectral index maps are
      \begin{equation}\label{eq:sigma pysm}
        \sigma_{\beta_{\rm D}}=0.04,\hspace{12pt}\sigma_{\beta_{\rm S}}= 0.22.
      \end{equation}
      It is worth noting that the level of realism of these simulations is similar to those used to quantify the performance of future $B$-mode facilities in e.g. \cite{2019JCAP...02..056A,2019arXiv190704473A,2020arXiv200812619T}.
      
      Three main aspects of these simulations that will challenge the assumptions made by our method are:
      \begin{itemize}
        \item The spectral index and amplitude maps are \emph{non-Gaussian}.
        \item The spectral index maps are based on existing observations, and therefore should be realistically correlated with the polarized amplitudes on large scales.
        \item The spectral index fluctuations are not necessarily distributed according to a power-law power spectrum on all scales.
      \end{itemize}

    \subsection{Instrumental effects}\label{ssec:sims.noise}
      All the simulations described in the previous two sections include instrumental noise designed to mimic an experiment following the specifications of the SO as described in \cite{2019JCAP...02..056A} (SO19 hereon). The simulations are generated in the six frequency bands covered by SO, centered around $\nu=27$, 39, 93, 145, 225 and 280 GHz. The three most relevant effects are instrument beam, scale-dependent noise, and inhomogeneous sky coverage. For simplicity, we do not include the effects of bandpass convolution and instead assume delta-function bandpasses centered at the frequencies listed above.
      
\begin{figure}
        \centering
        \includegraphics[width=0.7\textwidth]{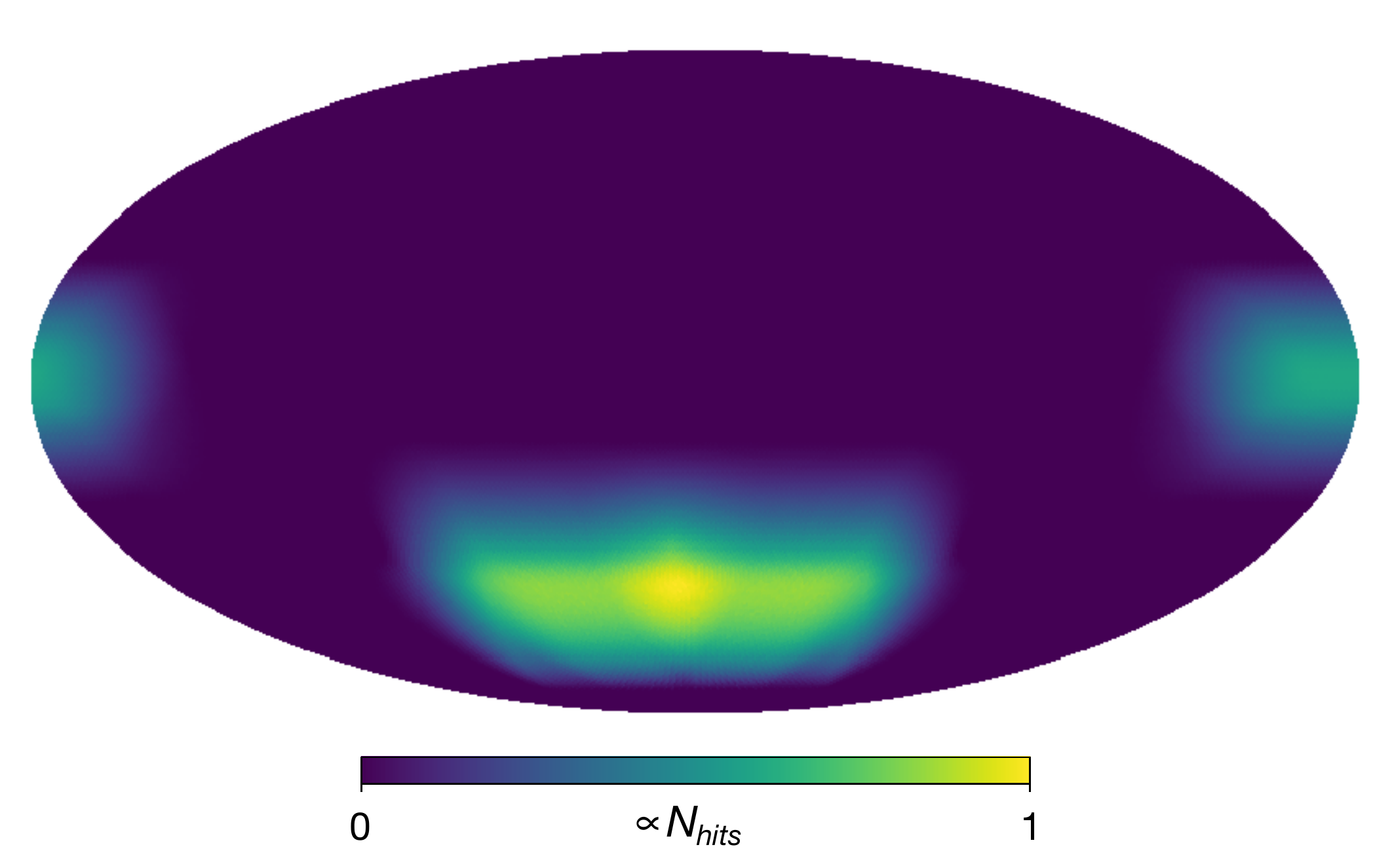}
        \caption{Sky mask used in the analysis, proportional to the map of hit counts in Equatorial coordinates used for the SO SATs.}\label{fig:nhits_mask}
      \end{figure}

      We generate noise realizations following the same two-step process described in SO19. First, we generate noise power spectra $N_\ell$ using the noise calculator released with the data supplement of SO19, assuming the baseline noise level, an optimistic knee scale $\ell_{\rm knee}$, and a total of 5 years of observation. Details can be found in the SO19 paper and in Table \ref{tab:SO}. We then generate homogeneous noise maps for the six frequency channels as Gaussian realizations of these power spectra. Finally, we scale these maps inversely with the square-root of the hits count map included in the data supplement (and shown in Figure \ref{fig:nhits_mask}) to account for the inhomogeneous sky coverage.
        
      \begin{table}[tbp]
        \centering
        \begin{tabular}{|c|c|c|c|c|}
          \hline
          Frequency & FWHM & Noise (baseline) & $\ell_{\rm knee}$ & $\alpha_{\rm knee}$\\
          (GHz) & (arcmin) & ($\mu K$-arcmin) & -- & --\\
          \hline
          27 & 91 & 35 & 15 & -2.4\\
          39 & 63 & 21 & 15 & -2.4\\
          93 & 30 & 2.6 & 25 & -2.5\\
          145 & 17 & 3.3 & 25 & -3.0\\
          225 & 11 & 6.3 & 35 & -3.0\\
          280 & 9 & 16 & 40 & -3.0\\
          \hline
        \end{tabular}
        \caption{Summary of the beam Full Width at Half Maximum (FWHM) apertures, baseline and goal sensitivity levels for each band of the SO Small Aperture Telescope (SAT) from \cite{2019JCAP...02..056A}. The correlated noise power spectrum is parametrized as $N_\ell=N_{\rm white}[(\ell/\ell_{\rm knee})^{\alpha_{\rm knee}}+1]$.}\label{tab:SO}
      \end{table}
         
      The signal maps are convolved with a Gaussian beam with a Full Width at Half Maximum (FWHM) aperture given by $\sigma_{\rm FWHM} = 1.22\lambda / D$, assuming diffaction-limited optics. Here $D=42\,{\rm cm}$ is the diameter of the SO Small Aperture Telescopes, $\lambda = \nu / c $ is the central wavelength of the band. Note that, for each simulation, we generate four independent noise realizations, each with a noise amplitude twice as large as the five-year SO sensitivity. This allows us to use these four realizations as independent data splits when estimating power spectra in order to avoid modeling the noise bias as part of the likelihood (see Section~\ref{ssec:theory.likelihood}). Signal and noise maps are then added in the observed footprint and saved to file.

    \subsection{Power spectrum measurement}\label{ssec:sims.cl}
      We extract the full set of multi-frequency $B$-mode power spectra $C_\ell^{\rm BB}$ from each simulation as follows.
      
      Each simulation consists of 24 pairs of $(Q,U)$ maps, corresponding to the 6 frequency channels and 4 data splits. We compute the cross-spectrum between all pairs of maps using a pseudo-$C_\ell$ estimator as implemented in {\tt NaMaster} \cite{2019MNRAS.484.4127A}. We use a differentiable sky mask constructed by smoothing the hits map provided with the SO19 data supplement with a $1^\circ$ FWHM beam and applying a ``C1'' apodization with a $5^\circ$ width (see \cite{2019JCAP...02..056A}) to the resulting map. The simulated map of hit counts in Equatorial coordinates for the SATs is displayed in Figure \ref{fig:nhits_mask}.
      
      Once all unique auto- and cross-spectra between the 24 different maps have been calculated, we produce coadded power spectra for every pair of frequencies $(\nu_1,\nu_2)$ by averaging over all power spectra involving maps at those frequencies corresponding to different data splits. By doing this, we isolate the contribution from the inhomogeneous noise bias to the auto-correlations, which are not used to generate the final coadded spectra. Discarding the auto-correlations leads to some loss of sensitivity. However, we find this loss to be negligible when comparing the final constraints on $r$ (presented in the next section) with the official SO forecasts \cite{2019JCAP...02..056A}. All power spectra were measured in a set of equi-spaced bandpowers with width $\Delta\ell=10$. Of these, only the 27 bandpowers in the range $30<\ell<300$ were used in the likelihood analysis, making the total size of the data vector $N_{\rm data} = 567$.
      
      As noted in Section \ref{ssec:sims.gaussian}, we repeat this process for two suites of 500 Gaussian simulations to generate the covariance matrices used in the likelihood analysis. We find each block of the resulting covariance involving two pairs of frequencies to be strongly dominated by its diagonal elements. Therefore, to reduce the statistical noise in the covariance due to the finite number of simulations, we set all off-diagonal elements in each block to zero, except for the diagonal and the first three superdiagonals.
      
      Any residual mode-coupling effects in the pseudo-$C_\ell$ estimator are taken into account analytically when evaluating the theory predictions for these measured power spectra \cite{2019MNRAS.484.4127A}.

  \section{Results}\label{sec:results}
    \subsection{Convergence of the model}\label{ssec:convergence}
\begin{figure}
        \centering
        \includegraphics[width=0.8\textwidth]{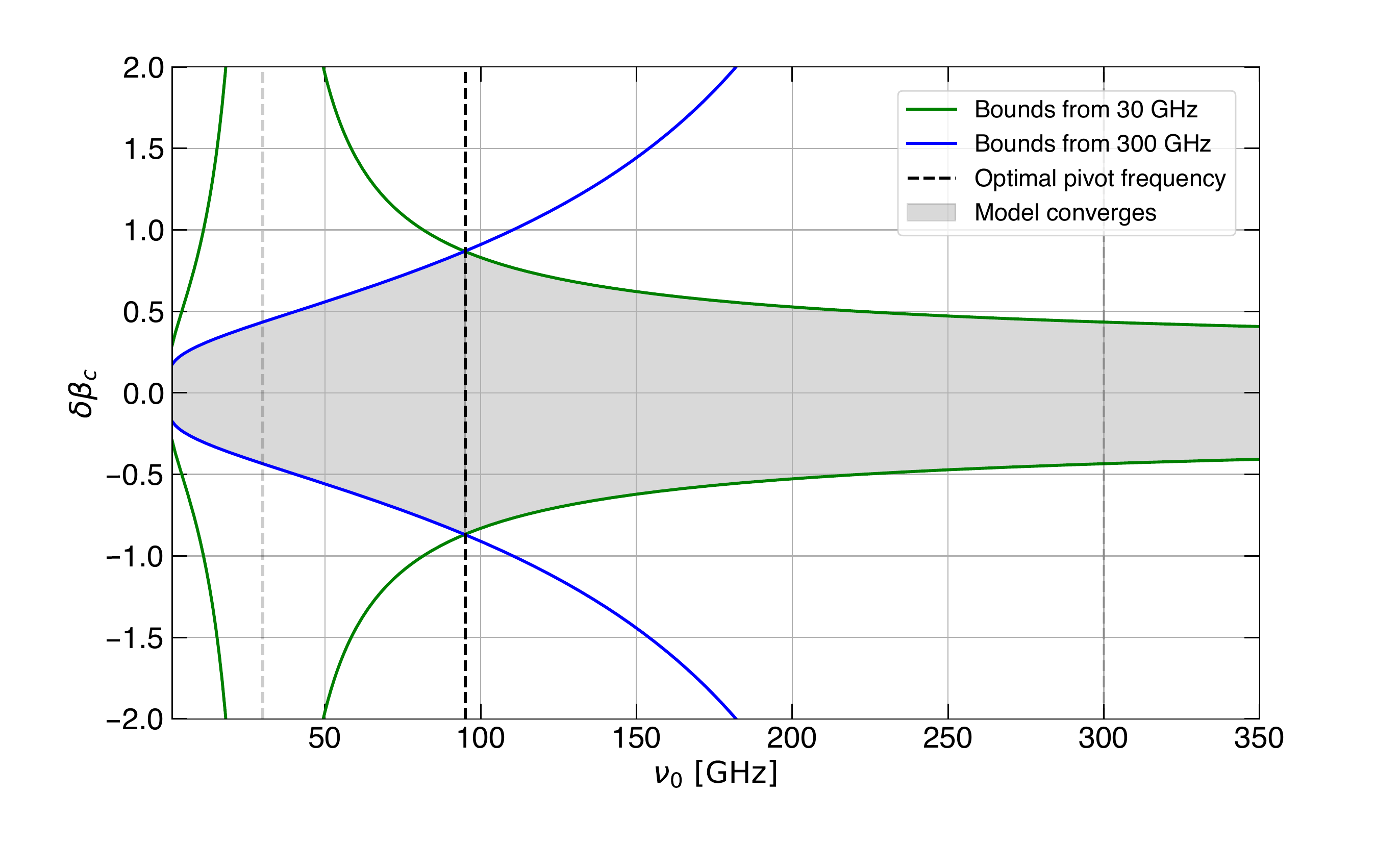}
        \caption{Convergence regions of the moment expansion method.}\label{fig:convergence}
      \end{figure}

      Before testing the method on simulated data, let us first gain some intuition on its behavior by studying the convergence of the power-spectrum-level moment expansion. The moment expansion will converge if each higher-order term in the series is monotonically decreasing~\cite{2017MNRAS.472.1195C}. Using the formalism described in Section~\ref{ssec:theory.map}, and assuming that all spectral parameters are spectral indices, Eq. \ref{eq:map_2} becomes a geometric series of the form
      \begin{equation}
        S^c_\nu(\beta(\nv)) =  \bar{S}^c_\nu\left(1 + x + x^2 + \dots \right)
      \end{equation}
      where $x = \log\left(\nu/\nu^c_0\right)\delta\beta_c$. The moment expansion will therefore converge as long as $x\lesssim1$. Thus, if we want the method to converge on the frequency interval $[\nu_1, \nu_2]$, there is a certain maximum $\delta\beta$ that the expansion can tolerate. Figure \ref{fig:convergence} shows the convergence bound on $\delta\beta$ for $\nu_1=30\,{\rm GHz}$ (blue) and $\nu_2=300\,{\rm GHz}$ (red) as a function of the pivot frequency $\nu_0$. The maximum $\delta\beta$ for the full region is achieved at $\nu_0=\sqrt{\nu_1\nu_2}\simeq95\,{\rm GHz}$ and corresponds to $|\delta\beta|_{\rm max} = 2/\log(\nu_2/\nu_1)\simeq0.87$.

      Thus, a good choice of pivot frequency can improve both the convergence of the model and minimize foreground-related biases for a given level of spectral index variability. Note, however, that the choice that maximizes the allowed variability of $\delta\beta$ for a converged expansion is not necessarily the optimal choice that minimizes foreground biases on $r$ for a finite order in the moment expansion. For example, taking the lowest-order expansion used here, and considering a model consisting of dust, synchrotron and CMB, the choice of pivot frequency for dust should balance the need to describe dust accurately in the frequency channels where it dominates the emission (280 and 220 GHz for SO), while providing a reasonable extrapolation of it on the CMB-sensitive frequencies (145 and 93 GHz for SO), regardless of the convergence of the model at all orders in the expansion in the furthest, synchrotron-dominated frequency channels (39 and 27 GHz for SO). We will study the impact of the choice of $\nu_0$ in Section \ref{ssec:res.gaussian}.

      Another convergence-related aspect of the particular model used here is the choice of power-law to describe the power-spectrum of spectral index variations $C_\ell^{\beta_c}$. Over a given range of scales $\ell_{\rm min}<\ell<\ell_{\rm max}$, the standard deviation of $\delta\beta_c$ is given by
      \begin{equation}
        \sigma^2(\beta_c)=\sum_{\ell=\ell_{\rm min}}^{\ell_{\rm max}}\frac{2\ell+1}{4\pi}C_\ell^{\beta_c}=\frac{B_c}{4\pi\ell_0^{\gamma_c}}\left[2\zeta(-\gamma_c-1)+\zeta(-\gamma_c)-3\right],
      \end{equation}
      where, in the second equality, $\zeta$ is the Riemann ``zeta'' function, we have assumed the power-law model used here (Eq. \ref{eq:cl_beta}), and used $\ell_{\rm min}=2$, $\ell_{\rm max}=\infty$. The standard deviation therefore diverges for $\gamma\geq-2$. The models used here, supported by current measurements of the synchrotron and dust spectral index \citep{2016A&A...594A...9P}, satisfy this constraint. However, as before, ultimately we only need this lowest-order expansion to describe the data on a limited range of scales, in which case these convergence constraints need not be strictly imposed.

    \subsection{Gaussian simulations}\label{ssec:res.gaussian}
      \begin{figure}
        \centering
        \includegraphics[width=1.\textwidth]{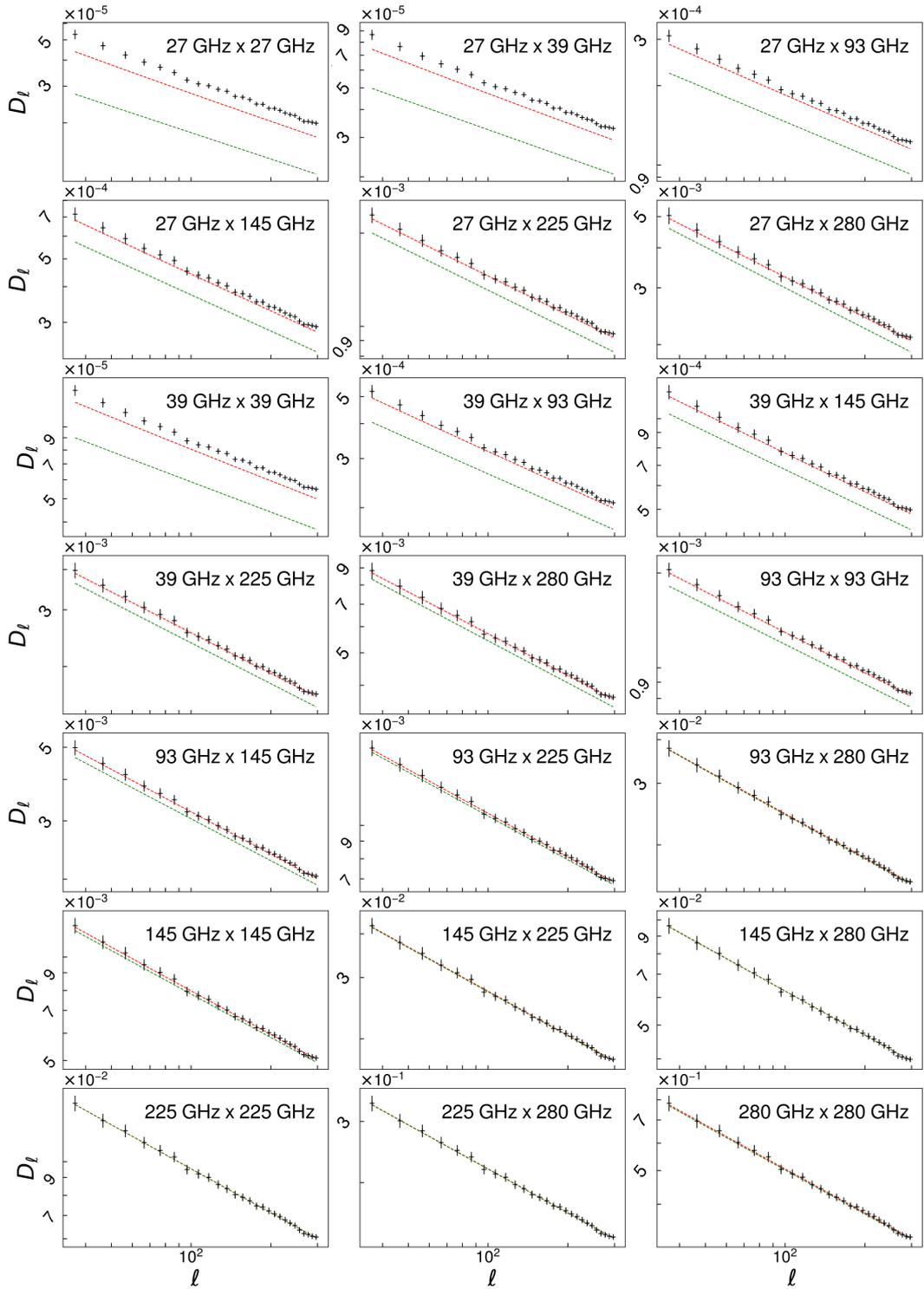}
        \caption{Power spectra of the average of ten Gaussian dust-only simulations, assuming a spatially varying spectral index with $\sigma(\beta_{\rm D})=0.3$ (\textit{black}). The lines show the prediction without accounting for this spatial variation ({\sl green}), and with the inclusion the moment expansion terms (\textit{red}). The order-0 prediction clearly underestimates the power spectrum on relevant frequencies (e.g. 93$\times$93 GHz), likely leading to a bias on $r$. The inclusion of the order-2 terms recovers the simulated data accurately up to effective frequencies $\sqrt{\nu\nu'}\simeq60\,{\rm GHz}$. At the lowest frequencies $(27,\,39\,{\rm GHz})$, where dust is subdominant to synchrotron and CMB emission, the order-2 expansion underpredicts the power spectrum significantly.}\label{fig:moments_added}
      \end{figure}
      We evaluate the performance of our method with a set of Gaussian simulations generated following the prescriptions described in Section~\ref{ssec:sims.gaussian}. We will study this performance as a function of the level of spectral index variation, parametrized by the standard deviation $\sigma_\beta$ rather than the amplitude parameters $B_c$.

      \subsubsection{Power spectrum predictions}
        First, we use these simulations to quantify the ability of the lowest-order moment expansionto describe Gaussian simulated data containing all of the higher order terms. The relevance of these terms will determine the bias on $r$ associated with the method, which should increase with increasing $\sigma_\beta$. Figure~\ref{fig:moments_added} displays the multi-frequency power spectra spectra $C_\ell^{\nu\nu'}$ in the range $30 \leq \ell \leq 300$ used here. The spectral index variation is $\sigma({\beta_D})=0.3$ with $\gamma_D = -3.5$ at $\nu_0 = 220$~GHz. These simulations contained only dust, the dominant foreground source for SO, with input parameters $A_{\rm D} = 5~\mu{\rm K}^2$, $\alpha_{\rm D} = -0.42$, $\beta_{\rm D} = 1.6$, in order to study the recovery of this particular component across the whole frequency range. Here the green line shows the predicted dust spectrum assuming a homogeneous spectral index, equal to the true mean $\beta_{\rm D}$ used in the simulations. The red line then shows the result of adding the $1\times1$ and $0\times2$ terms.
        
        We can see that the order-0 prediction clearly underestimates the power spectrum on relevant frequencies (such as the 93$\times$93 GHz combination), likely leading to a bias on $r$. However, in this particular example, considering only dust, the inclusion of the order-2 terms is able to recover the simulated data accurately up to effective frequencies $\sqrt{\nu\nu'}\simeq60\,{\rm GHz}$. At the lowest frequencies ($27,\,39\,{\rm GHz})$, where the emission is dominated by synchrotron, the order-2 expansion again underpredicts the power spectrum noticeably.
        
        \begin{figure}[!b]
          \centering
          \includegraphics[width=0.8\textwidth]{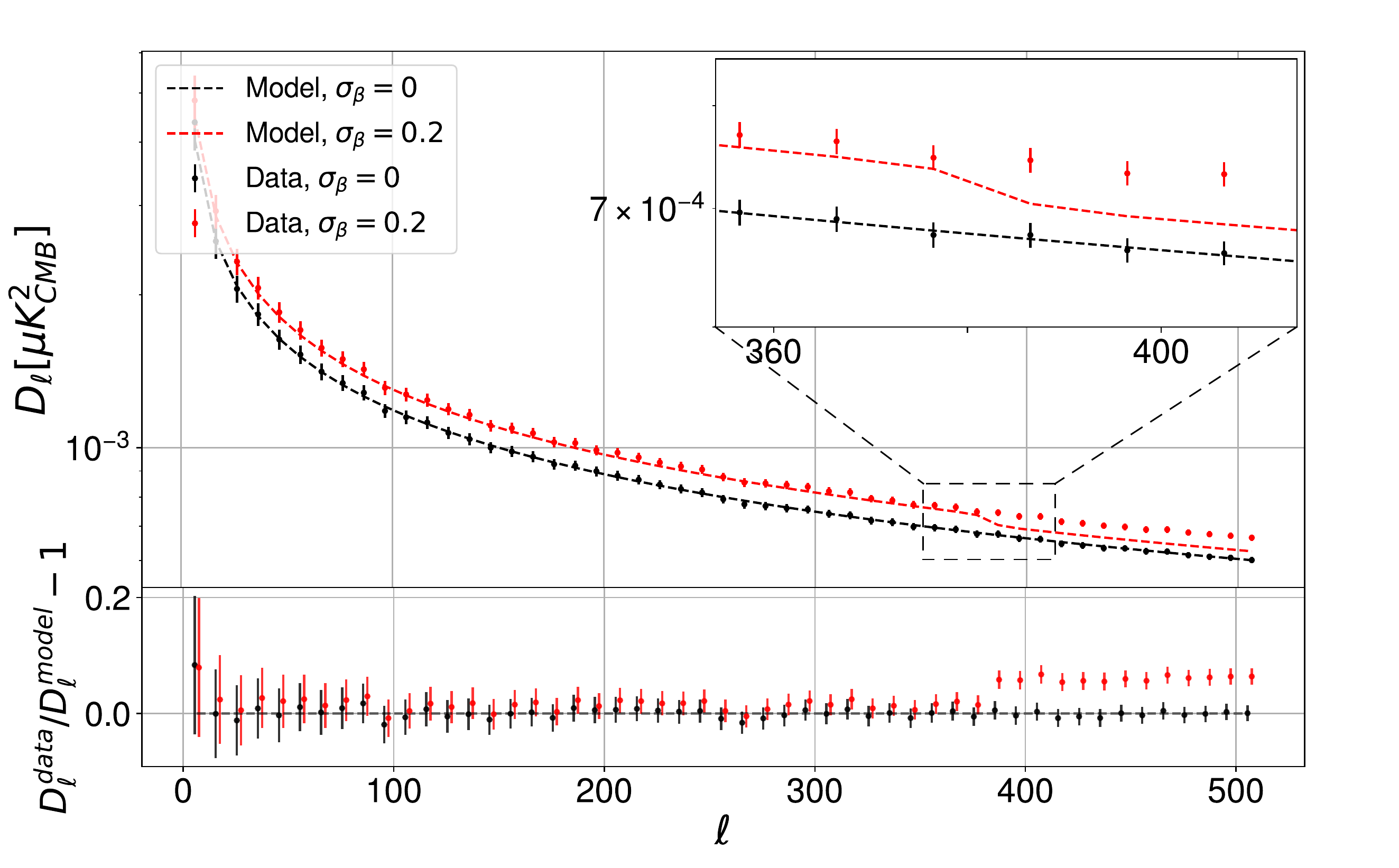}
          \caption{\textit{Upper panel:} Dust power spectrum at 93 GHz. The \textit{red} and \textit{black} points show the average of 10 Gaussian simulations with $\sigma_{\beta_{\rm D}}=0$ and 0.2 respectively. The \textit{dashed red} line shows the theoretical prediction with the moment expansion method using a maximum multipole $\ell_{\rm max}=384$ when convolving $C_\ell^{cc}$ and $C_\ell^{\beta_c}$ in the $1\times1$ term (Eq. \ref{eq:mom11}). As shown in the inset, the impact of this choice is limited to scales $\ell\gtrsim\ell_{\rm max}$, and thus this value must simply be chosen to lie outside the range of scales used in the analysis. \textit{Lower panel:} Fractional difference between the model and the simulated power spectra. The black error bars are recovered from the 10 realizations.} \label{fig:model_vs_sim_cls_all_comp}
        \end{figure}

        The $1\times1$ term in the moment expansion (Eq. \ref{eq:mom11}) involves a convolution of $C_\ell^{cc}$ and $C_\ell^{\beta_c}$. The associated sum over multipoles should in principle cover all integer $\ell_{1,2}$, however, in practice, it is only evaluated up to a maximum multipole $\ell_{\rm max}$. Figure \ref{fig:model_vs_sim_cls_all_comp} shows, as a dashed red line, the predicted dust auto-spectrum at $\nu=93\,{\rm GHz}$ using $\ell_{\rm max}=384$, compared with the average of ten Gaussian simulations with $\sigma_{\beta_{\rm D}}=0$ (black points) and $\sigma_{\beta_{\rm S}}=0.2$ (red points). The figure shows clearly that the moment expansion calculation becomes numerically inaccurate on scales $\ell>\ell_{\rm max}$, showing that the impact of the choice of $\ell_{\rm max}$ is limited to multipoles similar or larger than that scale. Thus, when implementing the moment expansion, it is sufficient to choose $\ell_{\rm max}$ to be slightly larger than the maximum multipole used in the analysis.

      \subsubsection{Constraints on $r$}      
        In order to propagate these results to final constraints on the tensor-to-scalar ratio, we proceed with the multi-frequency power spectrum likelihood analysis described in Section~\ref{ssec:theory.likelihood}. We start by producing a suite of 500 Gaussian simulations with constant spectral indices and $A_{\rm D}=5\,\mu{\rm K}^2, \ \alpha_{\rm D}=-0.42, \ A_{\rm S}=2\,\mu{\rm K}^2, \ \alpha_{\rm S}=-0.6$ as described in Section \ref{ssec:sims.gaussian}. These simulations are used throughout this analysis to compute the power spectrum covariance matrix and to validate our method. We also used these simulations to validate our pipeline and implementation of the \cite{2008PhRvD..77j3013H} likelihood, by making sure that we are able to recover the input foreground and CMB parameters in a subset of the suite.

\begin{figure}
          \centering
          \includegraphics[width=0.8\textwidth]{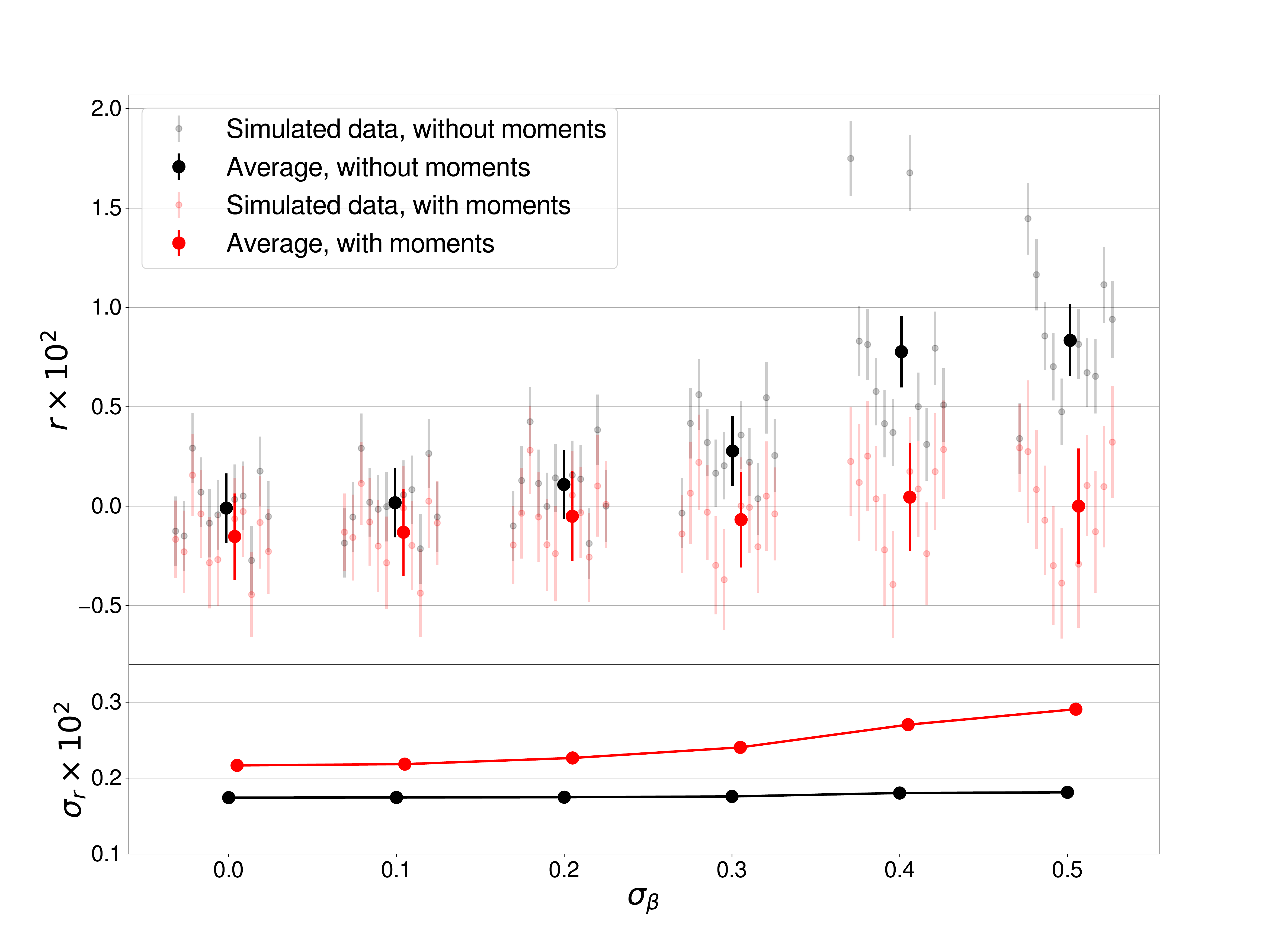}
          \caption{\textit{Upper panel:} Best-fit values of \textit{r} for ten realizations of the sky calculated at different values of $\sigma_{\beta_c}$ (the same for synchrotron and dust). Results are shown for a model assuming constant spectral indices (\textit{black}) and using the moment expansion method (\textit{red}). The position of each simulation in the $x$ axis is shifted slightly from its true $\sigma_{\beta_c}$ for clarity. The larger, solid dots, at the centre of each $\sigma_{\beta_c}$ value show the mean and standard deviation of each suite of simulations. \textit{Lower panel:} Statistical uncertainty $\sigma_r$ averaged over the ten realizations in the case of constant spectral indices (\textit{black}) and using the moment method (\textit{red}). The moment expansion is able to correct the bias on $r$ for all values of $\sigma_{\beta_c}$ considered, at the cost of increased final uncertainties with respect to a model with constant spectral indices (which themselves increase monotonically with $\sigma_{\beta_c}$).}\label{fig:r_correction}
        \end{figure}

        In the presence of spatially-varying spectral indices, a significant bias in the tensor-to-scalar ratio can arise if the corresponding effect in the multi-frequency spectra is not taken into account in the model. In the case of Gaussian simulated data, higher $B_c$ values generally correspond to higher biases on $r$. This is illustrated in Figure \ref{fig:r_correction}, which shows the recovered best-fit values of $r$ and their $1\sigma$ uncertainty for simulations run with increasingly larger values of $\sigma_\beta$ (equal for both synchrotron and dust). For each value of $\sigma_\beta$, a set of 10 simulations were generated, with the same seeds in each set, and the figure in the upper panel shows the individual results from each simulation as well as the average over simulations to minimize the impact of sample variance. All simulations were run with an input $r=0$, and therefore the bias on $r$ is directly given by its mean measured value. Results are shown for the final constraints on $r$ found assuming constant spectral indices (black dots) and using the moment expansion method to account for their spatial variation (red dots). We see in the bottom panel of \ref{fig:r_correction} that the statistical uncertainty on $r$ consistently increases with higher values of $\sigma_{\beta_c}$ when accounting for spectral index variation. The upper panel figure shows that the bias on $r$ when ignoring the spatial variation of $\beta_c$ grows with $\sigma_{\beta_c}$, becoming of the same order as the statistical uncertainties ($\sigma(r)\simeq0.002$) for $\sigma_{\beta_c}\sim0.25$. Accounting for the spectral index variation through our minimal moment expansion consistently corrects this bias, making it compatible with zero for the full range of $\sigma_{\beta_c}$ studied here, which encompasses the range of variation allowed by current data \citep{2016A&A...594A..10P}.

        It is worth noting that the mean of the posterior distribution of $r$ for data with no (or mild) spectral index variation analysed using the moment expansion is consistently biased low by $\Delta r\simeq-\sigma(r)\simeq-0.002$. This is due to the parameter degeneracy between $r$ and the spectral index amplitudes $B_c$, coupled with the positivity prior $B_c\geq0$ imposed on the latter. A positive $B_c$ increases the amplitude of the corresponding component's power spectra in the central, CMB-sensitive frequencies (93 and 145 GHz), which the model can compensate through a slightly negative $r$. We find, however, that the best-fit $r$ value reported here is not significantly biased. This is a well-known effect that arises also when parametrizing spectral index variation through a frequency decorrelation parameter $\Delta_c$ when imposing a physical prior $\Delta_c\leq1$ (see Appendix F of \cite{2018PhRvL.121v1301B}). The impact of physically-motivated priors should therefore be considered when interpreting the results of these analyses. We leave a more detailed study of this effect for future work.

        \begin{figure}
          \centering
          \includegraphics[width=0.8\textwidth]{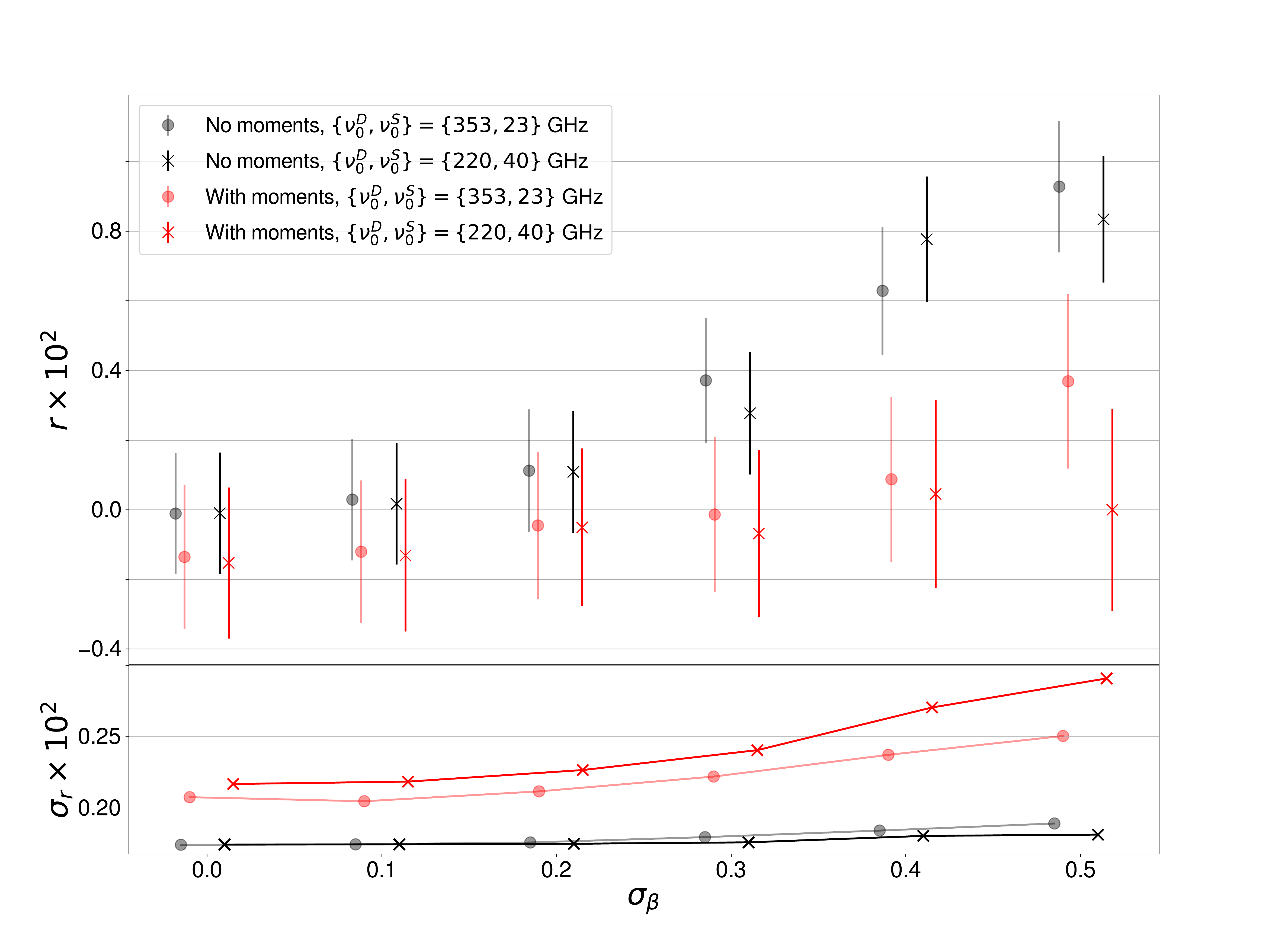}
          \caption{\textit{Upper panel:} Mean best-fit values of \textit{r} for ten sky simulations with varying levels of spectral index variation parametrized by $\sigma_{\beta}$, without accounting for moments (\textit{black}) and with moments correction (\textit{red}). Results are shown when the analysis is done using $40\,{\rm GHz}$ and $220\,{\rm GHz}$ as pivot frequencies for synchrotron and dust respectively ({\sl crosses}), and when using $23\,{\rm GHz}$ and $353\,{\rm GHz}$ ({\sl shaded circles}). The simulations were generated using the latter values as pivot frequencies for the foreground amplitude maps. Using pivot frequencies that coincide with channels close to the experiment's central bands allows the moment expansion to recover unbiased results on $r$ for the whole range of $\sigma_{\beta_c}$, whereas the alternative pivot frequencies lead to a bias for large $\sigma_{\beta_c}$. \textit{Bottom panel:} Statistical uncertainty on $r$ in the same cases.}\label{fig:r_corr_1sim}
        \end{figure} 

        The previous results were found with a model using pivot frequencies $\nu_0^{\rm D}=220\,{\rm GHz}$ and $\nu_0^{\rm S}=40\,{\rm GHz}$. This choice of pivot frequencies is motivated by the fact that SO has two pairs of foreground monitor channels at $27/39\,{\rm GHz}$ and $220/285\,{\rm GHz}$ for synchrotron and dust respectively. Since the allowed variability of $\beta_c$ is larger when the pivot frequencies are closer to the centre frequencies of the experiment's band, choosing the pivot frequencies to lie in the corresponding monitor channel lying closer to the CMB-sensitive bands ensures an accurate description of the foregrounds at the monitor frequencies and a reasonable extrapolation in the intermediate bands. An application of this method to instruments with more frequency bands, such as future space-borne missions \cite{hazumi2019litebird, 2020JLTP..199.1107S}, will likely require a more careful analysis of the optimal pivots. The relevance of this choice is shown in Figure~\ref{fig:r_corr_1sim}, where we compare the results from our previous analysis with the same method using pivot frequencies at $\nu_0^{\rm D}=353\,{\rm GHz}$ and $\nu_0^{\rm S}=23\,{\rm GHz}$, motivated by the location of sensitive bands in previous experiments \cite{2013ApJS..208...20B, 2016A&A...594A...1P}. When using these more distant pivot frequencies, we find that the moment expansion is not able to recover unbiased results for the highes value of $\sigma_{\beta_c}$ explored here.
       
        It is worth noting that the results above were obtained for simulations in which the spectral tilt of the spectral index fluctuations were $\gamma_{\rm D}=\gamma_{\rm S}=-3$. This corresponds to fairly steep power spectra, which induce substantial large-scale variations of the spectral indices. Since we analyse cut-sky simulations, the effective mean spectral indices in our sky patch are not necessarily centred at the fiducial values ($\bar{\beta}_{\rm D}=1.6$, $\bar{\beta}_{\rm S}=-3$) we assumed to generate the simulated data. In fact, due to the large scale power associated with $\delta\beta_c$, the variation of the effective mean spectral index in the analysed footprint can easily be as large as the $\sigma_\beta$ assumed in our simulations. In order to avoid this, we use large Gaussian priors on $\beta_{\rm D}$ and $\beta_{\rm S}$ (0.5 and 0.6 respectively). We have verified that the results presented here hold also for simulations with flatter spectral tilts $\gamma_c=-2.1$, in line with current measurements of the synchrotron spectral index \cite{2018A&A...618A.166K}.
        
        \begin{figure}[h!]
          \centering
          \includegraphics[width=0.48\textwidth]{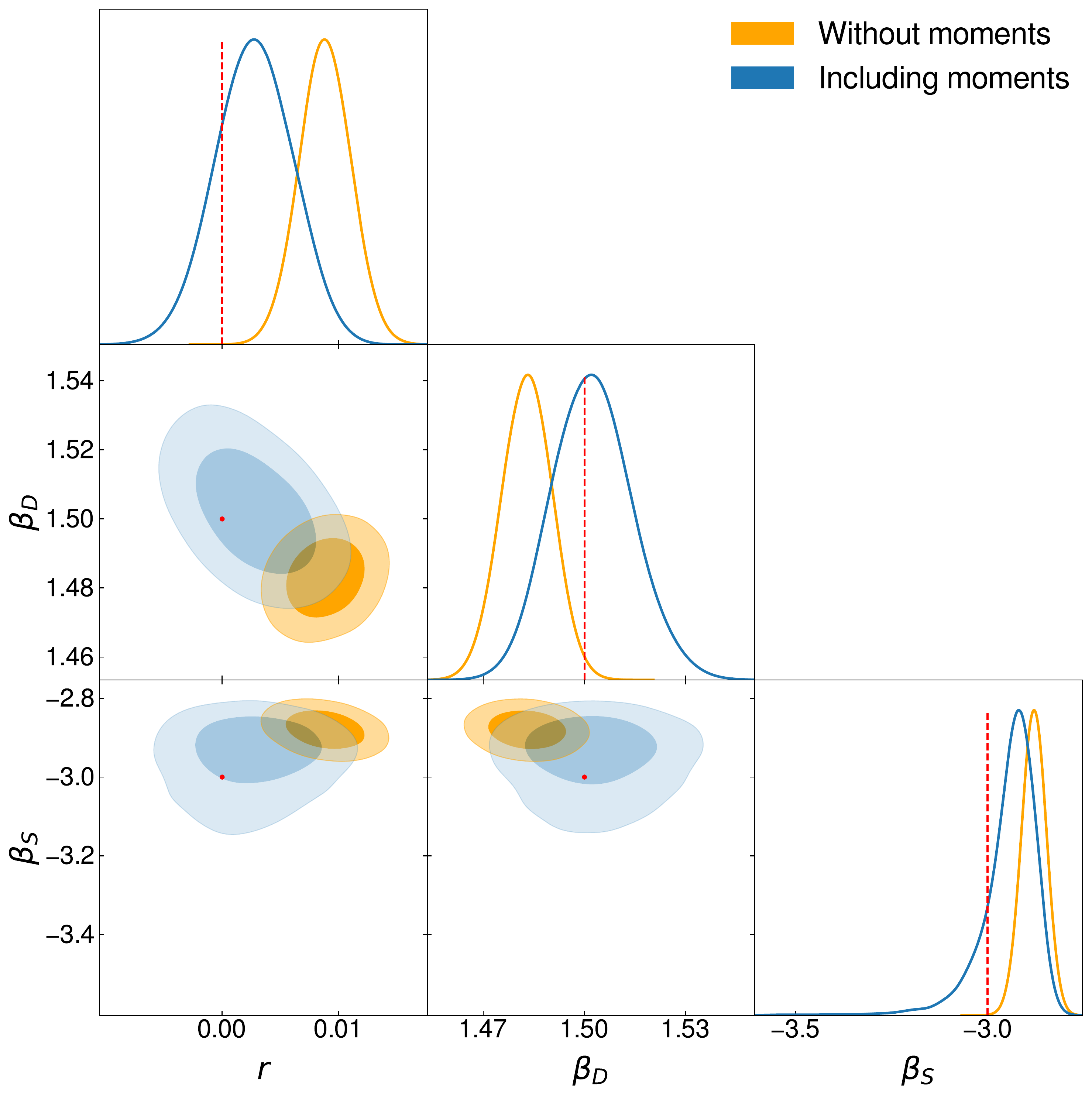}
          \includegraphics[width=0.48\textwidth]{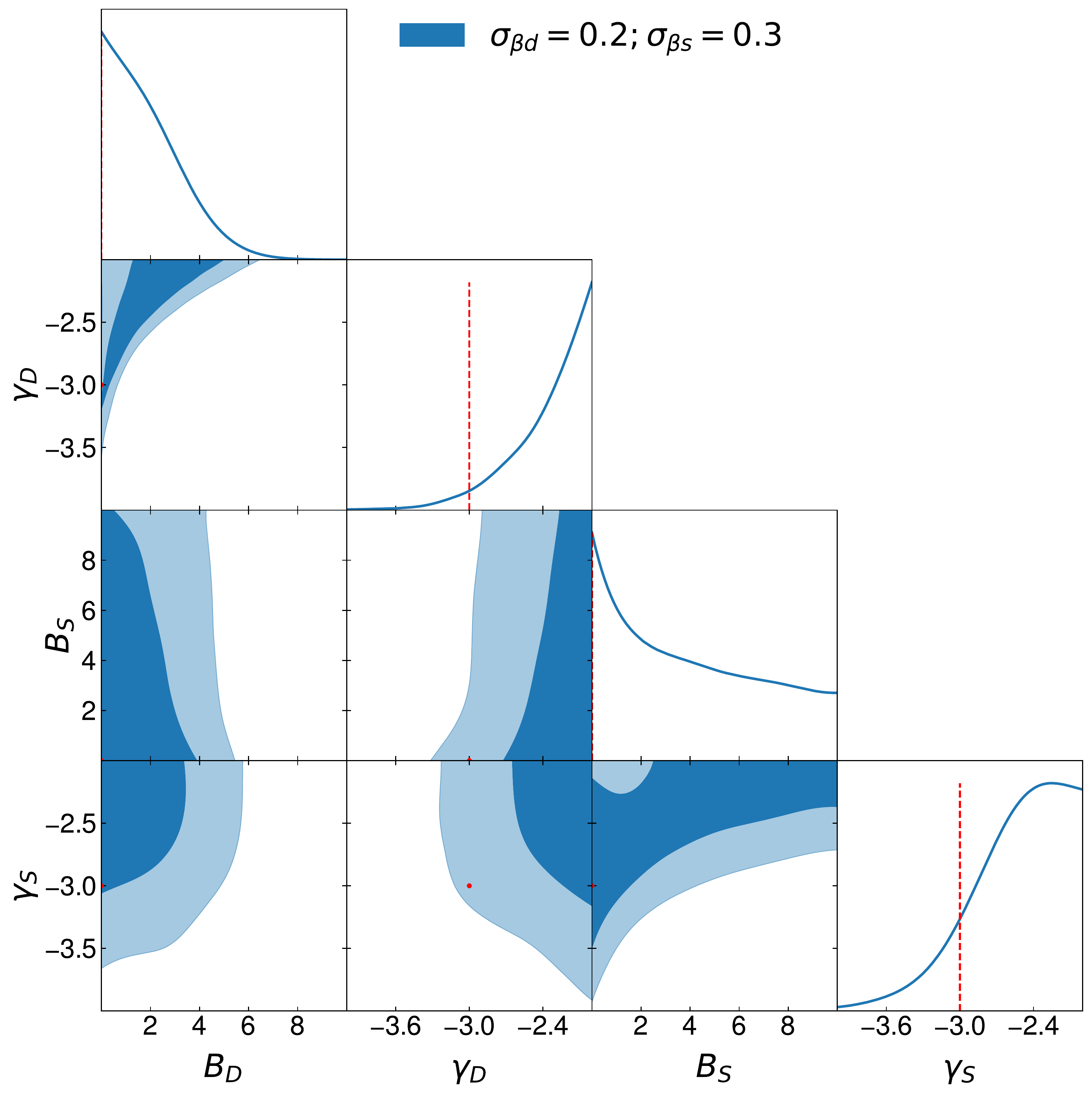}
          \caption{\textit{Left panel:} posterior distribution for $r$, $\beta_{\rm D}$ and $\beta_{\rm S}$ in Gaussian simulations with spatially-varying indices ($\sigma_{\beta_{\rm D}}=0.2$ and $\sigma_{\beta_{\rm S}}=0.3$). Results are shown for a model assuming constant spectral indices (\textit{orange} contours), and using the moment expansion method (\textit{blue} contours). The input parameters used to run the simulation are shown as red points and dashed lines. \textit{Right panel:} posterior distribution for the moment expansion parameters.} \label{fig:contour_gaus}
        \end{figure}
        
        The posterior distributions for one realization of a Gaussian simulation with $\sigma_{\beta_{\rm D}} = 0.2$, $\sigma_{\beta_{\rm S}} = 0.3$, $A_{\rm D} = 28 \mu{\rm K}^2$, $A_{\rm S} = 1.6 \mu{\rm K}^2$ is shown in Figure \ref{fig:contour_gaus}. The left panel shows the results for the tensor-to-scalar ratio and the average spectral indices with and without the inclusion of the additional moments parameters in blue and orange, respectively. The posterior distribution of the moment amplitudes and spectral tilts, displayed in the right panel of Figure \ref{fig:contour_gaus}, shows that $B_{\rm D}$ and $B_{\rm S}$ are not significantly larger than zero compared with their error bars. The power spectrum-based approach used here is therefore unable to detect the spectral index variation from the Gaussian simulated data, with a level of spectral index variation compatible with existing measurements of $\beta_{\rm S}$ and $\beta_{\rm D}$.

    \subsection{Realistic simulations}\label{ssec:res.pysm}
      Having validated the method for simulations that follow the implemented theoretical model, we now turn to simulations containing ``realistic'' foreground spectral index maps and amplitudes that do not adhere to our model assumptions (Gaussian fields, uncorrelated indices and amplitudes), as described in Section \ref{ssec:sims.pysm}. As discussed in Section~\ref{ssec:sims.pysm}, although the level of complexity increases compared to the Gaussian simulations, these simulations contain a comparatively low level of spectral index variation for both dust and synchrotron.
      \begin{figure}
        \centering
        \includegraphics[height=7.cm]{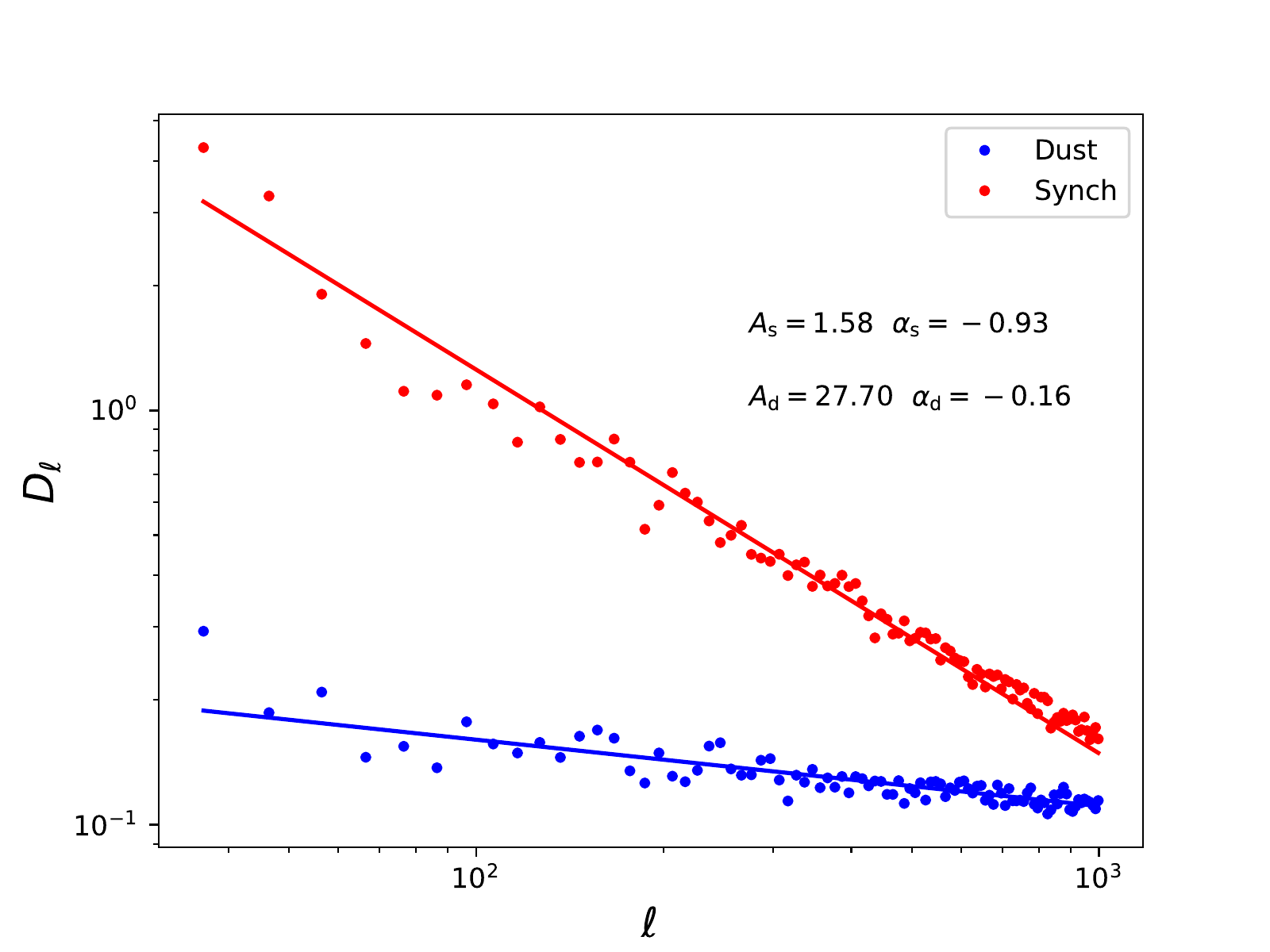}
        \caption{Power spectra of the \texttt{PySM} polarized synchrotron and dust maps (\textit{red} and \textit{blue} dots respectively) in the sky patch used in our analysis. The solid lines show the best-fit power-law spectra with amplitudes $(A_{\rm D},\,A_{\rm S}) =$ (27.70, 1.58)$\mu{\rm K}^2$ and tilts $(\alpha_{\rm D},\,\alpha_{\rm S}) =$ (-0.16, -0.93).}
        \label{fig:amplitudes_model}
      \end{figure}
        
      In Figure~\ref{fig:amplitudes_model} the power spectrum of the ``realistic'' amplitude maps for dust and synchrotron is compared with power-law fits. We find that, within the range of scales considered here, the power spectra of the {\tt PySM} maps are well described by power laws with amplitudes $(A_{\rm D},\,A_{\rm S}) =$ (27.7, 1.6)$\mu{\rm K}^2$ and tilts $(\alpha_{\rm D},\,\alpha_{\rm S}) =$ (-0.16, -0.93). Thus, any bias on $r$ resulting from the analysis of these simulations can be attributed to the spatially-varying spectral indices, and not to an incorrect modeling of the scale dependence of foreground amplitudes. We use these values to generate another suite of 500 Gaussian simulations (described in Section \ref{ssec:sims.gaussian}), which we use to estimate the power spectrum covariance. It is worth noting that the foreground power spectra exhibit clear departures from a perfect power law on scales larger than those used here \cite{2016A&A...594A..10P}, but this does not affect our results.
      \begin{figure}
        \centering
        \includegraphics[width=0.48\textwidth]{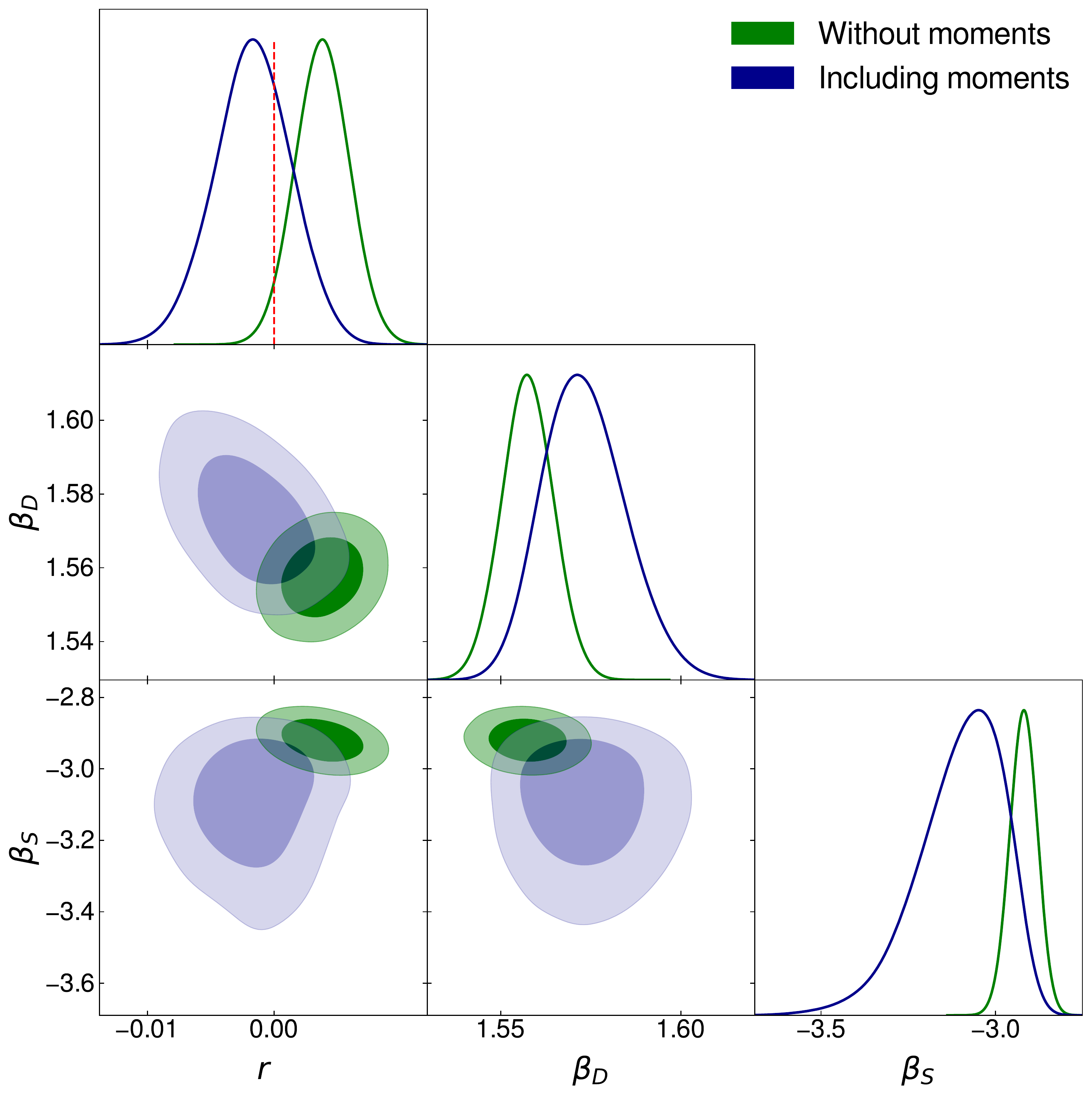}
        \includegraphics[width=0.48\textwidth]{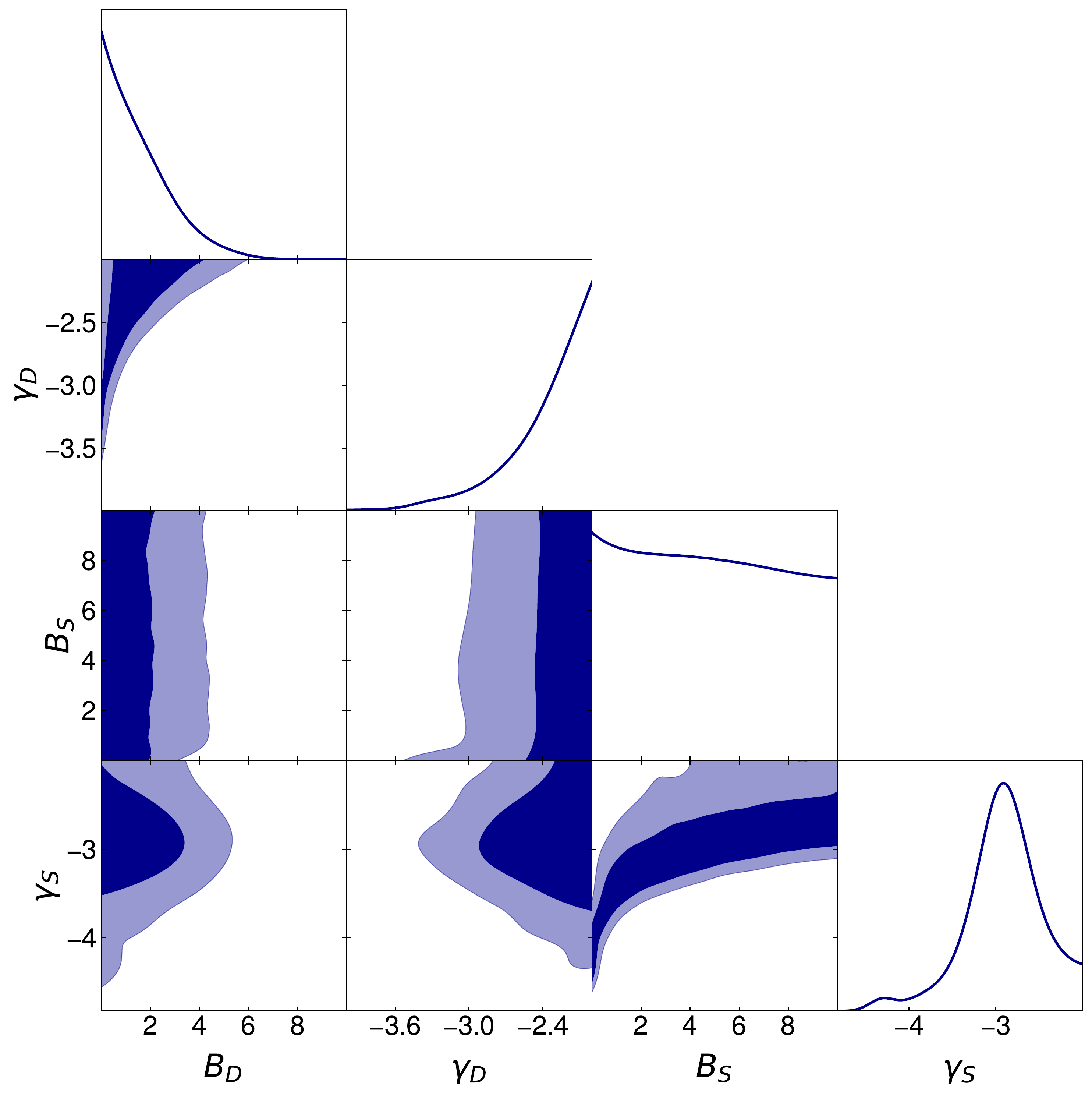}
        \caption{\textit{Left panel:} posterior distribution of the multi-frequency power spectrum likelihood for $r$, $\beta_{\rm D}$ and $\beta_{\rm S}$ from the realistic simulations. The green contours show the baseline result assuming constant spectral indices, while the blue curves show the constraints using the moment expansion method. \textit{Right panel}: distribution of the moment expansion parameters. No significant detection of spatially-varying indices is found with this method.}\label{fig:pysm}
      \end{figure}
        
      Ten different realizations of these realistic simulations were generated using the same foreground amplitude and spectral index maps, but varying the CMB and noise components. These were then analysed through our pipeline with and without the inclusion of the additional moments parameters to recover the posterior distribution for $r$. The results for one realization are shown in Figure~\ref{fig:pysm}. The baseline result of the $r$ posterior distribution is centered at 0.0029 $\pm$ 0.0023 averaged over the ten realizations. The higher complexity of these simulations introduces a bias on $r$, which is comparable to the bias found with the Gaussian simulations with a spectral index variation $\sigma_\beta = 0.3$. 
        
      After including the higher-order foreground terms and marginalizing over the 4 additional parameters, the constraints on $r$ are $r=0.0005 \pm 0.0028$ averaged over simulations. Thus, the bias on $r$ is corrected by over $1\sigma$ if we include the extra foregrounds parameters (the posterior of these is shown in the right panel of Figure~\ref{fig:pysm}). The inclusion of moments induces a small increase in the uncertainty on $r$, with $\sigma(r)=0.0028\pm0.0004$ averaged over simulations. This is visible in the corresponding wider posterior distribution of $r$ in Fig.~\ref{fig:pysm}. The spectral index parameters, on the other hand, seem to absorb some of the additional parameter freedom, with $\beta_{\rm D}$ increasing its posterior standard deviation by $\sim30\%$.

      \subsection{Simulation challenge}\label{ssec:res.challenge}
        As a final validation test for our implementation of the the moment expansion method, we have carried out a ``simulation challenge'' to determine the ability of the method to absorb a variety of foreground parametrizations. A set of 12 different simulations were generated independently by one of the authors (DA) and then analysed by a different author (SA) without knowledge of the contents of each simulation. In order to quantify the performance of the method, the analysis was carried out with and without the moment expansion. Covariance matrices were estimated using the second suite of Gaussian simulations described in Section \ref{ssec:sims.gaussian}. The simulations combined the levels of complexity encoded in the Gaussian and realistic simulations described in the previous sections with additional ingredients.
        \begin{itemize}
          \item Simulations were generated with Gaussian amplitudes, with $A_{\rm D}=28\,\mu{\rm K}^2$, $\alpha_{\rm D}=-0.16$, $A_{\rm S}=1.6\,\mu{\rm K}^2$, $\alpha_{\rm S}=-0.93$ (labelled ``G'' in Table \ref{tab:challenge}) and with the realistic amplitude templates included in {\tt PySM} (labelled ``P'' in Table \ref{tab:challenge}).
          \item The thermal dust contribution was propagated in frequency using the modified black-body spectrum in Eq.~\ref{eq:mbb} (labelled ``MBB'') as well as the model by \cite{2017ApJ...834..134H} (labelled ``H\&D'').
          \item Foreground spectral indices were generated as Gaussian fields (labelled by their standard deviation $\sigma_\beta$), as well as using the more complex templates described in Section \ref{ssec:res.pysm}.
          \item The underlying value of $r$ was varied between $r=0$ and $r=0.01$ in different simulations.
          \item Finally, two simulations were run using the statistical model described in \cite{2017A&A...603A..62V} (labelled ``VS'' in the table). In this model, the three-dimensional structure of the GMF is described by a finite number of layers (we use $N_{\rm layer}=7$ layers). The coherent component of the magnetic field is the same in all layers, while its turbulent part is generated as a Gaussian random field. Once the direction of the GMF is determined, maps of the $Q$ and $U$ Stokes parameters are generated by scaling the dust intensity map found by \planck{} \cite{2016A&A...594A..10P}. As an additional level of complexity, and in an attempt to describe the three-dimensional distribution of the dust spectral index, we associate each layer with a different Gaussian realization of $\delta\beta_{\rm D}$ with standard deviation $\sigma_{\beta_{\rm D}}=0.13$ (the combined rms variation for 7 layers is $\sigma_{\beta_{\rm D}}\simeq0.35$).
        \end{itemize}
        \begin{table}[tbp]
          \centering
          \begin{tabular}{|l|c|c|c|c|c|}
            \hline
            \multicolumn{2}{|c|}{Simulation} & \multicolumn{2}{c|}{No moments} & \multicolumn{2}{c|}{With moments}\\
            \hline
            Description; ($\sigma_{\beta_{\rm D}},\sigma_{\beta_{\rm S}}$) & $r_{\rm true}$ & $r_{\rm fit}\pm\sigma_r$ & $\chi^2/{\rm d.o.f.}$ & $r_{\rm fit}\pm\sigma_r$ & $\chi^2/{\rm d.o.f.}$\\
            \hline
            G, MBB; $\sigma_\beta=(0,0)$ & 0 & -0.0013 $\pm$ 0.0021 & 0.8 & -0.0024 $\pm$ 0.0024 & 0.8 \\
            G, MBB; $\sigma_\beta=(0,0)$ & 0.01 & 0.0116 $\pm$ 0.0022 & 0.8 & 0.0099 $\pm$ 0.0025 & 0.8 \\
            \hline
            G, MBB; $\sigma_\beta=(0.2,0.3)$ & 0 & \textcolor{red}{0.0088 $\pm$ 0.0023} & 0.9 & 0.0038 $\pm$ 0.0035 & 0.8 \\
            G, MBB; $\sigma_\beta=(0.2,0.3)$ & 0.01 & \textcolor{red}{0.0158 $\pm$ 0.0025} & 0.9 & 0.0098 $\pm$ 0.0035 & 0.9 \\
            \hline
            P, MBB; $\sigma_\beta={\tt PySM}$ & 0 & \textcolor{red}{0.0051 $\pm$ 0.0022} & 0.9 & 0.0036 $\pm$ 0.0026 & 0.9 \\
            P, MBB; $\sigma_\beta={\tt PySM}$ & 0.01 & 0.0130 $\pm$ 0.0023 & 0.9 & 0.0104 $\pm$ 0.0027 & 0.9 \\
            \hline
            G, H\&D; $\sigma_\beta=(0,0)$ & 0 & \textcolor{red}{0.0058 $\pm$ 0.0026} & 1.1 & 0.0003 $\pm$ 0.0037 & 1.1 \\
            G, H\&D, $\sigma_\beta=(0,0)$ & 0.01 & 0.0122 $\pm$ 0.0024 & 1.1 & 0.0055 $\pm$ 0.0038 & 1.1 \\
            \hline
            P, H\&D; $\sigma_\beta={\tt PySM}$ & 0 & \textcolor{red}{0.0052 $\pm$ 0.0025} & 1.1 & 0.0001 $\pm$ 0.0033 & 1.1 \\
            P, H\&D; $\sigma_\beta={\tt PySM}$ & 0.01 & 0.0120 $\pm$ 0.0024 & 1.1 & 0.0069 $\pm$ 0.0034 & 1.1 \\
            \hline
            P, VS; $\sigma_\beta=(0.13,N.A.)$ & 0 & \textcolor{red}{0.0114$\pm$ 0.0024} & 1.0 & -0.0036 $\pm$ 0.0036 & 1.0 \\
            P, VS; $\sigma_\beta=(0.13,N.A.)$ & 0.01 & \textcolor{red}{0.0184 $\pm$ 0.0025} & 1.0 & \textcolor{red}{0.0029 $\pm$ 0.0034} & 1.0 \\
            \hline
          \end{tabular}
          \caption{Results from the simulation channel. The values marked in red show the results with a bias $|r_{\rm fit}-r_{\rm true}|\geq 2 \sigma_r$. }\label{tab:challenge}
        \end{table}        

        The results for the different simulations run as part of this challenge are summarized in Table \ref{tab:challenge}. Highlighted in red are the results of simulations in which a the best-fit value of $r$ was found to be more than $2\sigma$ away from the input value. We find that in most cases where foregrounds introduce a bias on $r$ at this level using the standard method, the moment expansion method is able to recover unbiased results at the same level.

        The results of the analysis for the first four simulations (rows 1-4 in Table \ref{tab:challenge}) are consistent with the results presented in Section \ref{ssec:res.gaussian}. When no spectral index variation is introduced ($\sigma_\beta=(0,0)$), the final results both with and without moments are compatible with the true values of $r$ within their respective $\sigma(r)$, with a modest widening in the final constraints when moments are included. When Gaussian spectral index variation is introduced ($\sigma_\beta=(0.2,0.3)$), the baseline result exhibits a bias up to the $3\sigma$ level, which is reduced by $2\sigma$ using the moment expansion method, with a $\sim30\%$ increase in uncertainty.

        Similarly, the $1-2\sigma_r$ level bias induced by the ${\tt PySM}$ templates for the spectral indices (rows 5 and 6) is reduced by $\gtrsim85\%$ of $\sigma(r)$ compared to the baseline result, with a $17\%$ degradation in $\sigma(r)$.

        When the H\&D dust SED is used in the simulations (rows 7 to 10), the deviations from the original MBB spectra introduce a bias at the 1-2$\sigma$ level. The additional freedom in the dust model due to the inclusion of the moment parameters corrects the bias whilst also widening the final constraints by $\sim40\%$.
        
        Finally, the more complex non-Gaussian VS simulations induce a large bias on $r$, which the moment expansion is able to correct at the cost of increasing the final uncertainties by $\sim50\%$. We find, however, that for an input $r=0.01$, the moments method underestimates it by $\simeq2.1\sigma$. After finding this, we verified that the same result is reproduced on a second realization of the VS model, and therefore believe that this is not due to a statistical fluke. The cause of this bias is not clear. It could be due to the additional complexity of the VS simulations (non-Gaussianity, additional spectral index variation and frequency decorrelation). It could also be that the simulated data in this case differs significantly from the model used to construct the covariance matrix and the resulting likelihood is ill behaved. We leave a more thorough analysis of the performance of the lowest-order moment expansion used here on data with this level of complexity for future work.

      \subsection{BICEP2/Keck Array data}
        \begin{figure}[t]
          \centering
          \includegraphics[width=.48\textwidth]{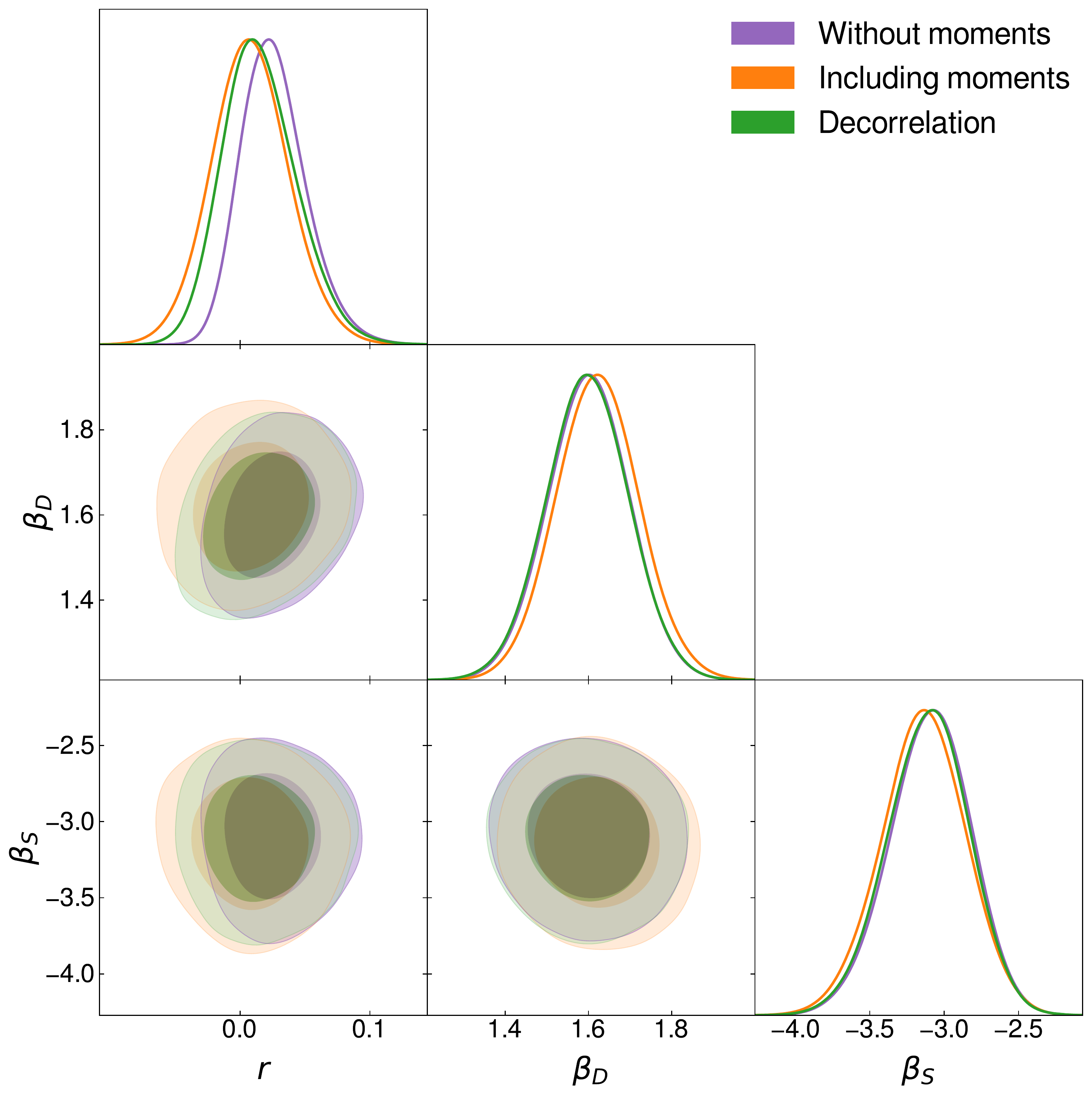}
          \includegraphics[width=.48\textwidth]{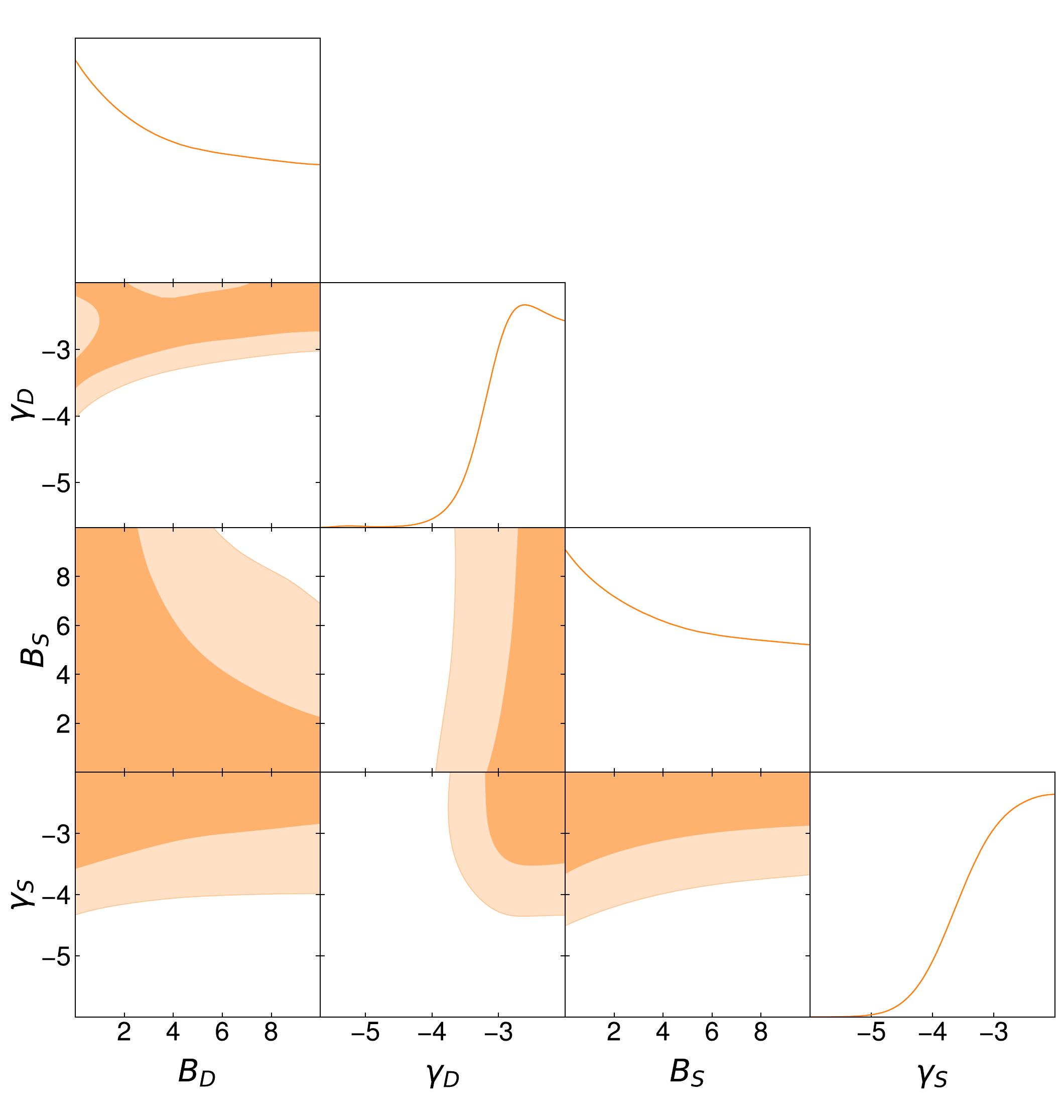}
          \caption{\textit{Left}: Posterior distribution of \textit{r}, and spectral indices $\beta_{\rm D}$ and $\beta_{\rm S}$ using our analysis pipeline on the publicly available BK15X data. The green curves show the baseline moments-less case and agrees with the BK15X published posterior distributions, with \textit{r} peaking at 0.023. Marginalizing over foregrounds spatial variation shifts the \textit{r} curve closer to zero (orange) without a significant impact on the posteriors. The results using the decorrelation parameter method, as done in \cite{2018PhRvL.121v1301B}, are displayed in purple. \textit{Right}: Distribution of the additional SED parameters corresponding to the moments of the parameter distributions.}\label{fig: BK15}
        \end{figure}
        In order to further validate the moment expansion method, as well as to explore the sensitivity of current $B$-mode constraints to the spatial variability of foreground spectral indices, we have applied the method to the latest publicly available data from the \bicep{} collaboration \cite{2018PhRvL.121v1301B} (BK15X hereon).
        
        The BK15X dataset is fully described in \cite{2018PhRvL.121v1301B}. The power spectrum data contain cross-correlations between 12 frequency bands, including 3 \bicep{} bands (at 95, 150 and 220 GHz), 7 \planck{} bands (30, 44, 70, 100, 143, 217 and 353 GHz) \cite{2016A&A...594A...1P} and 2 low-frequency WMAP bands (23 and 33 GHz) \cite{2013ApJS..208...20B}. The analysis is performed over the \bicep{} footprint, covering approximately 400 deg$^2$.  All the information concerning cross-correlations between different frequencies and polarization bands, their covariance matrix, all frequency bandpass transmission curves, and bandpower window functions are publicly available\footnote{See \url{http://bicepkeck.org/bk15_2018_release.html}.}. All power spectra have been measured in a set of 9 equi-spaced bandpowers which cover the multipole range $\ell\lesssim400$. The total size of the data vector is $N_{\rm data} = 2700$, of which 702 elements correspond to $B$-mode-only correlations.

        In order to validate the implementation of our multi-frequency foreground model, we first reproduced the fiducial BK15X results by running our component separation pipeline on the $B$-mode data using the same parameter priors used by BK15X. The results are displayed in green in Figure~\ref{fig: BK15}. We recover the published $r$ posterior distribution \cite{2018PhRvL.121v1301B} almost perfectly, and other foreground parameter constraints (e.g. on $\beta_{\rm S}$, $\beta_{\rm D}$) agree to better than 5\% with those presented in BK15X.
        
        The orange contours in the same figure show the constraints after marginalizing over the four additional moment expansion parameters. We observe a small shift in the $r$ posterior mean towards smaller values, accompanied by a broadening of the distribution by $\sim24\%$. The right panel of Figure \ref{fig: BK15} shows the posterior distribution of the moment amplitudes and spectral tilts. We find no evidence of spectral index variations in the BK15X data.
        
        The original BK15X analysis studied the impact of spatially-varying spectral indices by introducing two frequency decorrelation parameters \cite{2018PhRvL.121v1301B} with different levels of scale dependence. As we show in Appendix \ref{app:corr}, a constant decorrelation parameter is equivalent to a non-perturbative moment expansion in the specific case of scale-independent (i.e. uncorrelated) spectral index variations\footnote{Note that, although the analysis of BK15X allowed for different forms of scale dependence for the decorrelation parameter, these do not map directly onto a model for the spatial fluctuations in the foreground spectral indices. We have only considered the constant decorrelation case here.}. In order to compare the impact of both parametrizations on the final constraints on $r$, we reproduced the decorrelation results found by BK15X with our pipeline. The results, shown as purple contours in Fig. \ref{fig: BK15}, display a similar downward shift in the $r$ posterior, while its width is slightly smaller than the full moment expansion. Thus, although parametrizing the impact of spatially-varying indices in terms of frequency decorrelation captures one of the most important effects in the data vector, the moment expansion is able to effectively marginalize over additional freedom in the scale dependence of these spatial variations. Although the relevance of this additional freedom is small in current datasets, it may prove to be important when more sensitive data become available.

\section{Conclusion}\label{sec:conclusion}
    The potential of primordial $B$-modes to open a new window into the physics of the very early Universe makes the search for this faint CMB polarization signal one of the most compelling goals of modern cosmology. However, their detection is challenged by several analysis and technical challenges. On the analysis side, a detailed characterization of Galactic polarized emission is necessary to disentangle CMB $B$-modes from other sources of polarized emission. In this context, multi-frequency power-spectrum-based component separation pipelines have been used by ground-based experiments to derive the current state-of-the-art constraints on the tensor-to-scalar-ratio $r$ from $B$-modes \cite{2018PhRvL.121v1301B}. $C_\ell$-based methods provide several advantages when handling ground-based $B$-mode data, since they are computationally less challenging than pixel-based techniques, and allow for a straightforward treatment of correlated noise, complicated map filtering, and certain systematics \citep{2020arXiv201102449A}.
    
    This method, however, has major drawbacks when applied to data with higher sensitivity over wider patches of the sky. In particular, the characterization of spatially-varying foreground spectra, being difficult to model at the power spectrum level, poses major challenges. We have addressed this issue here by designing a power-spectrum-based component separation approach, based on existing moment expansion methods. In order to curb the number of new free parameters that rapidly appear in the standard series expansion, which would degrade the final constraints on $r$ significantly, we impose three strong assumptions on the model: spectral index variations are Gaussianly distributed, foreground amplitudes and spectral index variations are uncorrelated, and spectral index variations of different foreground sources are uncorrelated. Departures from these assumptions are thus ignored as higher-order terms in the expansion. The resulting parametric model has four additional free parameters: the amplitudes and slopes of the power spectra describing the dust and synchrotron spectral index fluctuations.

    In order to quantify the performance of this method, as well as the impact of its assumptions on the final $r$ constraints, we have made use of a suite of sky simulations with increasing degrees of realism. These include Gaussian foreground simulations following the same assumptions of the model (Section \ref{ssec:res.gaussian}), as well as more realistic simulations based on various models proposed in the literature (Sections \ref{ssec:res.pysm} and \ref{ssec:res.challenge}). These simulations assumed an instrumental setup similar to that expected of the SO Small-Aperture Telescopes ($\sim10\%$ of the sky with a $\sim2\mu{\rm K}\,{\rm arcmin}$ white-noise level), accounting for both inhomogeneous sky coverage and non-white noise. Finally, we have applied this method to the $B$-mode data made publicly available by the \bicep{} collaboration.
    
    Overall we find that the method is able to correct the bias to the tensor-to-scalar ratio induced by spectral index variations for most realistic foreground models. From the Gaussian simulation suite we find that the leading-order expansion is able to cope with spectral index variations at the level of $\sigma_\beta\lesssim0.5$, compatible with existing measurements of the polarized foreground spectral indices \cite{2016A&A...594A..10P,2018A&A...618A.166K}. We also find that a judicious choice of the pivot frequencies defining the amplitude and spectral indices for a given foreground source can improve the performance of the method. Although we do not attempt an exact derivation of the optimal pivots, we follow the rule of thumb of using a pivot frequency corresponding to the foreground monitor channel closest to the foreground minimum.
    
    The additional freedom in the foreground model due to the four moment parameters results in a moderate widening of the final constraints on $r$. In the case of an SO-like dataset, this corresponds to a $\sim30-50\%$ increase in $\sigma(r)$, with a more moderate increase of $\sim20\%$ for the current \bicep{} data. As shown in Appendix \ref{app:corr}, the moment expansion used here is a generalization of the frequency decorrelation parameter used in the BK15X analysis, and we find that both methods have a similar effect on the posterior distribution for those data. Although the levels of spectral index variability explored here can induce a bias on the measurement of $r$ at the level of $1$ or $2\sigma$, we find that it is not possible to significantly detect the effects of this variation on the foreground multi-frequency power spectra for SO-like sensitivities.

    The model used here is based on a leading-order expansion of the cross-frequency power spectra with respect to the spectral index variations. We show in Appendix \ref{app:corr} that the impact of spatially-varying indices on the power spectrum can be calculated exactly at all orders using methods developed in the context of CMB lensing reconstruction. Although the applicability of other lensing-inspired techniques in the context of non-Gaussian foregrounds may be limited, it could be an interesting avenue to pursue in the future.
    
    Given this option to include new foreground parameters, one natural question to ask is whether these parameters are necessary to describe a given set of real data. One could perform simple model selection tests using information criteria or use a evidence-based methods \cite{2009MNRAS.398.1601F} to judge whether moment parameters are required by the data. Given that the results here show only a fairly moderate increase in $\sigma_r$, we leave this analysis to future work. It should be noted that such an exploration is warranted even on the standard $B$-mode analyses in the literature (e.g. to study whether, from a model selection perspective, frequency decorrelation parameter needs to be included in the model \citep{2018PhRvL.121v1301B}).

    Our study has been limited to the analysis of primordial $B$-modes from ground-based facilities targeting the recombination bump on scales $30\lesssim\ell\lesssim300$. Its applicability to space missions targeting the reionization bump on larger scales, and covering a wider range of frequencies \cite{hazumi2019litebird,2014JLTP..176..733M}, may be be limited, and the use of pixel-based methods is likely more appropriate. Nevertheless, we expect that the methodology presented here, as well as its potential extensions, some already explored in the literature \citep{2019arXiv191209567M}, will be useful in the analysis of future ground-based observatories, such as the SO \cite{2019JCAP...02..056A} or CMB Stage-4 \cite{2016arXiv161002743A}, which will require the characterization of spatially-varying foreground spectra, and marginalization over them in order to achieve reliable constraints on $r$.

\acknowledgments
  We thank Erminia Calabrese, Jens Chluba, Josquin Errard and Aditya Rotti for many helpful discussions. SA is funded by a Kavli/IPMU PhD Studentship. MHA acknowledges support from the Beecroft Trust and Dennis Sciama Junior Research Fellowship at Wolfson College. This project has received funding from the European Research Council (ERC) under the European Union’s Horizon 2020 research and innovation programme (grant agreement No.~693024). DA acknowledges support from the Beecroft Trust, and from the Science and Technology Facilities Council through an Ernest Rutherford Fellowship, grant reference ST/P004474. TM and NK acknowledge the World Premier International Research Center Initiative (WPI), MEXT, Japan for support through Kavli IPMU.

\appendix
\section{Non-perturbative calculation and CMB lensing}\label{app:corr}
  If the spectral properties of a given component were perfectly homogeneous, the maps of that component at different frequencies would be simply rescaled versions of the same field. The presence of spatially-varying spectral parameters, however, causes additional perturbations on this field at different frequencies. The situation is similar to the basic description of the lensed CMB: an unperturbed field (in this case the primordial CMB fluctuation) is perturbed by a non-linear modification caused by another field (the lensing deflection). Furthermore, the minimal model explored here, in which both foreground amplitudes and spectral index fluctuations are treated as uncorrelated Gaussian random fields, makes the analogy between both phenomena almost exact. Because of this, we can use some of the methods developed within the context of CMB lensing to improve on the moment expansion method used in this paper. In particular, this appendix presents a full, non-perturbative calculation of the multi-frequency power spectrum for a given component in analogy to the calculation of the lensed CMB power spectrum.

  Let us start by considering a single component $c$ with amplitude $T_c(\nv)$ at a pivot frequency $\nu_0$, and a spectrum of the form $S_\nu^c(\beta_c)=\left(\nu/\nu_0\right)^{\beta_c}F_\nu^c$,
  where $F_{\nu}$ is an arbitrary function of frequency normalized to $F_{\nu_0}^c=1$, and $\beta_c(\nv)=\bar{\beta}_c+\delta\beta_c(\nv)$ is the component's spatially-varying spectral index with mean $\bar{\beta}_c$. At a frequency $\nu$, the component's sky emission is
  \begin{equation}
    T_{c,\nu}(\nv)=S_\nu^c(\beta_c(\nv))\,T_c(\nv)=\bar{S}_\nu^c\,T_c(\nv)\,e^{x_\nu\,\delta\beta(\nv)},
  \end{equation}
  where $\bar{S}_\nu^c\equiv S_\nu^c(\bar{\beta}_c)$, and we have defined $x_\nu\equiv\log(\nu/\nu_0)$.

  The multi-frequency correlation function of the perturbed field is defined as
  \begin{equation}\label{eq:corr1}
    \xi^{\nu\nu'}(\theta)\equiv\left\langle T_{c,\nu}(\nv)\,T_{c,\nu'}(\nv')\right\rangle=\bar{S}_\nu^c\,\bar{S}_{\nu'}^c\left\langle T_c(\nv)\,T_c(\nv')\right\rangle\,\left\langle e^{x_\nu\,\delta\beta(\nv)+x_{\nu'}\,\delta\beta(\nv')}\right\rangle,
  \end{equation}
  where $\cos\theta\equiv\nv\cdot\nv'$ and, in the second equality, we have assumed that $T_c$ and $\delta\beta_c$ are statistically independent.

  The second expectation value can be calculated analytically using the following well-known result for Gaussian variables:
  \begin{equation}
    \langle e^y\rangle\equiv\int_{-\infty}^\infty dy\frac{e^{-y^2/(2S^2)}}{\sqrt{2\pi S^2}}e^y=e^{S^2/2},
  \end{equation}
  where $S^2$ is the variance of $y$. Applying this result to the last term in Eq. \ref{eq:corr1} we obtain
  \begin{equation}\label{eq:corr2}
    \xi^{\nu\nu'}(\theta)=\bar{S}_\nu^c\,\bar{S}_{\nu'}^c\,\xi^{cc}(\theta)\,\exp\left[(x_\nu^2+x_{\nu'}^2)\frac{\sigma_{\beta_c}^2}{2}+x_\nu x_{\nu'}\,\xi^\beta(\theta)\right],
  \end{equation}
  where we have defined the unperturbed correlation function and the spectral index correlation function:
  \begin{equation}
    \xi^{cc}(\theta)\equiv\left\langle T_c(\nv)T_c(\nv')\right\rangle,\hspace{12pt}
    \xi^\beta(\theta)\equiv\left\langle\delta\beta_c(\nv)\,\delta\beta_c(\nv')\right\rangle.
  \end{equation}
  Correlation functions and power spectra of scalar fields are related to each other through:
  \begin{align}\label{eq:cl2xi}
    &\xi^X(\theta)=\sum_{\ell=0}^\infty\frac{2\ell+1}{4\pi}L_\ell(\cos\theta)\,C^X_\ell\simeq \int_0^\infty\frac{d\ell\,\ell}{2\pi}J_0(\ell\theta)\,C_\ell^X,\\\label{eq:xi2cl}
    &C^X_\ell=2\pi \int_0^\pi d(\cos\theta)\,L_\ell(\cos\theta)\,\xi^X(\theta)\simeq2\pi\int_0^\infty d\theta\,\theta\,J_0(\ell\theta)\,\xi^X(\theta),
  \end{align}
  where the second equality in each line is valid in the flat-sky approximation.
  \begin{figure}
    \centering
    \includegraphics[width=0.7\textwidth]{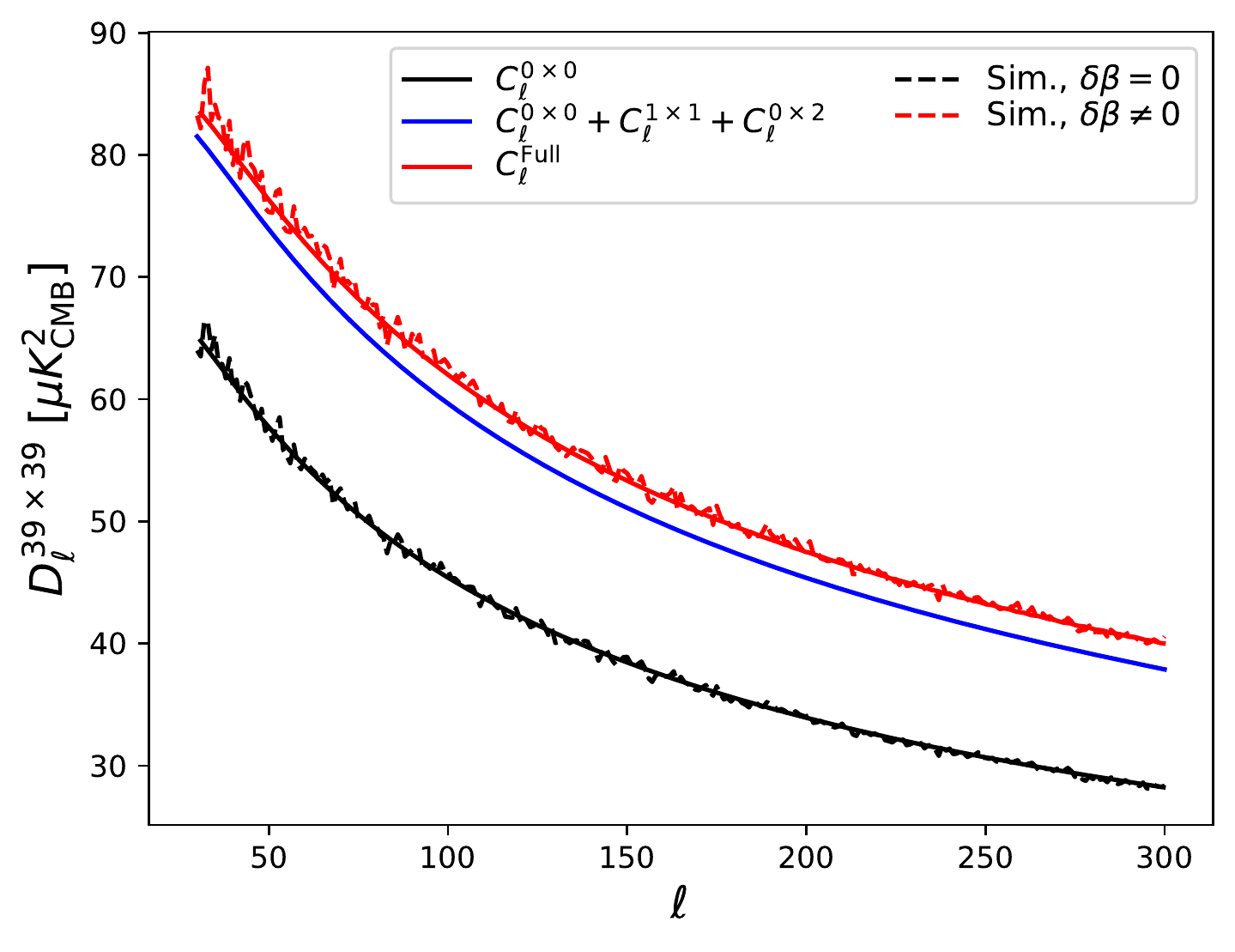}
    \caption{Average dust power spectrum at 39 GHz from 100 Gaussian simulations with spectral index variation $\sigma_{\beta_{\rm D}}=0$ (dashed black) and $\sigma_{\beta_{\rm D}}=0.16$ (dashed red) with a pivot $\nu_0^{\rm D}=353\,{\rm GHz}$. The solid black line shows the theoretical prediction assuming no spectral index variation, while the solid blue line shows the leading-order moment expansion used in this paper. The solid red line shows the exact calculation described in Equations \ref{eq:corr2}, \ref{eq:cl2xi} and \ref{eq:xi2cl}.}
    \label{fig:xi_cl}
  \end{figure}

  Given a model for the power spectra of the foreground amplitude and spectral index variations $C_\ell^{cc}$ and $C_\ell^{\beta_c}$, the multi-frequency power spectrum can be computed at all orders in $\delta\beta$ through a 3-step process:
  \begin{enumerate}
    \item Calculate $\xi^{cc}(\theta)$ and $\xi^\beta(\theta)$ (and $\sigma_\beta^2\equiv\xi^\beta(0)$) from $C_\ell^{cc}$ and $C_\ell^\beta$ using Eq. \ref{eq:cl2xi}. 
    \item Calculate $\xi^{\nu\nu'}(\theta)$ from $\xi^{cc}(\theta)$ and $\xi^\beta(\theta)$ using Eq. \ref{eq:corr2}.
    \item Calculate $C^{\nu\nu'}_\ell$ from $\xi^{\nu\nu'}(\theta)$ using Eq. \ref{eq:xi2cl}.
  \end{enumerate}
  The Hankel transforms translating between correlation functions and power spectra (Eqs. \ref{eq:cl2xi} and \ref{eq:xi2cl}) can be calculated using computationally efficient methods (e.g. \cite{Murray2019}).

  Figure \ref{fig:xi_cl} shows the improvement of this method over the lowest-order moment expansion (Eq. \ref{eq:c_ell_1}). The figure shows the dust power spectrum at 39~GHz for $\sigma_{\beta_{\rm D}}=0.16$ calculated from the average of 100 Gaussian simulations (dashed red), the zeroth-order approximation $C_\ell^{\nu\nu'}|_{0\times0}$ (black), the lowest-order moment expansion (Eq. \ref{eq:c_ell_1}, blue), and the full calculation described here (solid red). The non-perturbative calculation is able to recover the simulated power spectrum exactly.

  Two interesting limits can be explored in Eq. \ref{eq:corr2}. First, for small spectral index variations, the first term in the Taylor expansion of the exponential factor in this equation leads to two contributions, proportional to $\sigma_\beta^2$ and $\xi^\beta(\theta)$. These correspond to the real-space versions of the moment expansion terms $C_\ell^{\nu\nu'}|_{0\times2}$ and  $C_\ell^{\nu\nu'}|_{1\times1}$ respectively. Secondly, in the limit of uncorrelated spectral index variations ($C_\ell^{\beta_c}=\Omega_{\rm pix}\sigma_{\beta_c}^2$, where $\Omega_{\rm pix}$ is the pixel size), the frequency decorrelation is scale-independent, and given by \cite{2017A&A...603A..62V}
  \begin{equation}
    \frac{C_\ell^{\nu\nu'}}{\sqrt{C_\ell^{\nu\nu}C_\ell^{\nu'\nu'}}}=\exp\left[-\frac{1}{2}\sigma^2_{\beta_c}\Omega_{\rm pix}\log^2\left(\frac{\nu}{\nu'}\right)\right],
  \end{equation}
  recovering the parametrization in terms of decorrelation parameter used by \cite{2018PhRvL.121v1301B}.

  Even with the use of fast methods for Hankel transforms, implementing the full model slows down the computation of the likelihood used here significantly, and therefore all our results use the moment expansion. As we have shown in section \ref{ssec:res.gaussian}, the accuracy of the expansion is sufficient for the range of frequencies and sensitivities explored here. Furthermore, although the full calculation yields unbiased results, its applicability is fairly limited to the case of Gaussian spectral index variations in two dimensions that are statistically uncorrelated with the foreground amplitudes. The inaccuracies associated with these assumptions are likely to be more important than the differences with the moment expansion calculation. That being said, the intuition gained from the CMB lensing analysis could be useful for other applications, such as employing lensing reconstruction techniques to multi-frequency maps in order to recover spectral index maps (see e.g. \cite{2018MNRAS.479.5577P}).

  \section{Spin-$s$ generalization of the first-order expansion}\label{app:spin2}
    The expressions in Eqs. \ref{eq:mom00}, \ref{eq:mom11} and \ref{eq:mom02} can be easily generalized to the case of spin-$s$ quantities (as is the case for the spin-2 CMB polarization field). Using the same formalism and notation presented in \cite{2019MNRAS.484.4127A}, the map ${\bf m}_\nu$ and amplitudes ${\bf T}_c$ in Eq. \ref{eq:map_2} are promoted to spin-$s$ fields with both $Q$ and $U$ components in real space, and $E$ and $B$-mode components in harmonic space:
    \begin{align}\nonumber
      &m_\nu\hspace{6pt}\longrightarrow\hspace{6pt}{\bf m}_\nu\equiv(m^Q_\nu,m^U_\nu),\\\nonumber
      &a_{\ell m}\hspace{6pt}\longrightarrow\hspace{6pt}{\bf a}_{\ell m}\equiv(a^E_{\ell m},a^B_{\ell m}),
    \end{align}
    while the spectral index variations $\delta\beta_c$ remain scalar, real-valued fields. The power spectrum in Eq. \ref{eq:cldef} between any two spin-$s$ field then becomes a 2$\times$2 matrix containing the four correlations between their $E$ and $B$ components:
    \begin{equation}
      \langle {\bf a}_{\ell m}{\bf b}^\dagger_{\ell' m'}\rangle\equiv\delta_{\ell\ell'}\delta_{mm'}{\sf C}^{ab}_\ell\equiv\delta_{\ell\ell'}\delta_{mm'}\left(
      \begin{array}{cc}
           C_\ell^{a^Eb^E}& C_\ell^{a^Eb^B} \\
           C_\ell^{a^Bb^E}& C_\ell^{a^Bb^B} 
      \end{array}
      \right).
    \end{equation}
    
    Following the same techniques used in the derivation of the standard pseudo-$C_\ell$ estimator (see e.g. \cite{2019MNRAS.484.4127A}), it is easy to show that the expressions in Eqs. \ref{eq:mom00} and \ref{eq:mom02} for the $0\times0$ and $0\times2$ contributions remain formally unchanged:
    \begin{align}
      &{\sf C}_\ell^{\nu\nu'}|_{0\times0}=\bar{S}^{\rm D}_\nu\bar{S}^{\rm D}_{\nu'}\,{\sf C}_\ell^{\rm DD}+\bar{S}^{\rm S}_\nu\bar{S}^{\rm S}_{\nu'}\,{\sf C}_\ell^{\rm SS}+\left(\bar{S}^{\rm D}_\nu\bar{S}^{\rm S}_{\nu'}+\bar{S}^{\rm S}_\nu\bar{S}^{\rm D}_{\nu'}\right){\sf C}^{\rm SD}_\ell,\\
      &{\sf C}_\ell^{\nu\nu'}|_{0\times2}=\sum_{c\in\{{\rm D},{\rm S}\}}\frac{1}{2}\left[\bar{S}^c_\nu\,\partial^2_\beta\bar{S}^c_{\nu'}+\bar{S}^c_{\nu'}\,\partial^2_\beta\bar{S}^c_\nu\right]{\sf C}_\ell^{cc}\sigma_{\beta_c}^2,
    \end{align}
    while the $1\times1$ term in Eq. \ref{eq:mom11} becomes
    \begin{equation}
      {\sf C}_\ell^{\nu\nu'}|_{1\times1}=\sum_{c\in\{{\rm D},{\rm S}\}}\partial_\beta\bar{S}^c_\nu\,\partial_\beta\bar{S}^c_{\nu'}\sum_{\ell_1\ell_2}\frac{(2\ell_1+1)(2\ell_2+1)}{4\pi}\wtj{\ell}{\ell_1}{\ell_2}{s}{-s}{0}^2\,C_{\ell_2}^{\beta_c}\,\hat{\sf d}_{S_\ell}{\sf C}_{\ell_1}^{cc}\hat{\sf d}^\dagger_{S_\ell},
    \end{equation}
    where $S_\ell\equiv\ell+\ell_1+\ell_2$, and we have defined the matrix
    \begin{equation}
      \hat{\sf d}_n\equiv\frac{1}{2}\left(
      \begin{array}{cc}
         1+(-1)^n & -i[1-(-1)^n] \\
         i[1-(-1)^n] & 1+(-1)^n
      \end{array}
      \right).
    \end{equation}

    Thus we see that the main additional effect of the spectral index variation on the spin-$2$ polarized foregrounds is the mixing of $E$ and $B$ modes in the $1\times1$ term. This is similar to the generation of CMB lensing $B$ modes from primordial $E$ modes, following the analogy with CMB lensing used in Appendix \ref{app:corr}, or to the leakage between $E$ and $B$ in the presence of a sky mask. In the case of the featureless foreground power spectra considered here, this effect is degenerate with the unknown amplitude of the spectral index variations, and therefore we find the spin-$0$ approximation described in Section \ref{sec:theory} accurate enough for our main analysis, as demonstrated in Section \ref{ssec:res.gaussian}.

\bibliography{bibliography}

\end{document}